\def\hst{{\sl HST}}
\def\nicmos{{\sl NICMOS}}

\def\gc {{GC}}
\documentclass[12pt,preprint]{aastex}
\usepackage{epsf}
\usepackage{color}
\usepackage{graphicx}
\usepackage{graphics}
\usepackage{epsfig}
\usepackage{longtable}

\begin{document}
\title{HST Paschen-$\alpha$ Survey of the Galactic Center: Data
  Reduction and Products}

\author{H. Dong$^{1}$, Q. D. Wang$^1$, A. Cotera$^2$, S.
Stolovy$^3$, M. R. Morris$^4$, J. Mauerhan$^3$, E. A. Mills$^4$,
G. Schneider$^5$, D. Calzetti$^1$, C. Lang$^6$}

\affil{$^1$ Department of Astronomy, University of Massachusetts,
Amherst, MA 01003} \affil{$^2$ SETI Institute} \affil{$^3$ Spitzer
Science Center, California Institute of Technology} \affil{$^4$
Department of Physics and Astronomy, University of California, Los
Angeles}\affil{$^5$ Steward Observatory, University of Arizona}
\affil{$^6$ Department of Physics and Astronomy, University of
Iowa}
\affil{E-mail: hdong@astro.umass.edu, wqd@astro.umass.edu}

\begin{abstract}
Our \hst /\nicmos\ Paschen-$\alpha$ survey of the Galactic center,
first introduced in~\citet{wan10}, 
provides a uniform, panoramic, high-resolution map of stars and
ionized diffuse gas in the central $416 {\rm~arcmin^2}$ of
the Galaxy. This survey was carried out  with 144 \hst\ orbits using
two narrow-band filters at 1.87 ${\rm \mu m}$ and 1.90 ${\rm \mu m}$ in NICMOS
Camera 3. In this
paper, we describe in detail the data reduction and mosaicking
procedures followed, including background level matching and
astrometric corrections. We have detected $\sim$ 570,000 near-IR sources using the `Starfinder' 
software and are able to quantify
photometric uncertainties of the detections. 
The source detection limit
varies across the survey field but the typical 50\% completion limit
is $\sim 17$th mag (Vega System) in the 1.90 ${\rm \mu m}$ band. A comparison with the
expected stellar magnitude distribution shows that these sources
are primarily Main-Sequence massive stars (${\rm \gtrsim 7 M_\odot}$)
and evolved lower mass stars at the distance of the Galactic center. In
particular, the observed source magnitude distribution exhibits a
prominent peak, which could represent the Red Clump (RC) stars within
the Galactic center. The observed magnitude and color of these RC
stars support a steep extinction curve in the near-IR toward the
Galactic center~\citep{nis09}. The flux ratios of our detected sources in the two bands also allow for an
adaptive and statistical estimate of extinction across the field.
With the subtraction of the extinction-corrected continuum, we
construct a net Paschen-$\alpha$ emission map and identify a set
of Paschen-$\alpha$-emitting sources, which should mostly be
evolved massive stars with strong stellar winds. The majority of the 
identified Paschen-$\alpha$ point sources are located within the 
three known massive Galactic center stellar clusters. However, a significant 
fraction of our Paschen-$\alpha$-emitting
sources are located outside the clusters and may represent a new
class of `field' massive stars, many of which may have formed in
isolation and/or in small groups. The maps and source catalogues
presented here are available electronically.
\end{abstract}

\section{Introduction}\label{s:introduction}
Our Galactic center~\citep[GC; 1$^{\prime\prime}$=0.04 ${\rm pc}$ at our adopted distance
of $\sim$8.0 ${\rm kpc}$,][]{ghe08,gil09,rei09} is the only galactic
nuclear region where stellar population can
be resolved. The GC thus provides an unparalleled opportunity
to understand the star formation (SF) mode and history
under an extreme environment, characterized by high
temperature, density, turbulent velocity, and magnetic field of the
interstellar medium (ISM), in addition to strong gravitational
tidal force (e.g., Morris \& Serabyn 1996).

In HST/GO 11120, we have carried out the first large-scale, high spatial resolution,
near-IR survey of the GC over a field of $\sim 39^\prime \times
15^\prime$ around Sgr A$^*$ (or 93.6 ${\rm pc}$ $\times$ 36 ${\rm pc}$ at the GC
distance). Using the NICMOS Camera 3 (NIC 3) aboard the Hubble Space
Telescope (HST), providing an instrumental spatial resolution of
$\sim$ 0.2$\arcsec$ ($\sim$ 0.01 ${\rm pc}$ at the distance of the GC), our survey spatially
resolves more than 80\% of the light detected in the 1\% wide F187N
(Paschen-$\alpha$, 1.87 ${\rm \mu m}$) and F190N (adjacent continuum, 1.90 $\mu
m$) bands.In our previous paper~\citep{wan10}, we
have presented an overview of the survey, including its rationale
and design as well as a brief description of the data reduction,
preliminary results, and potential scientific implications. The
survey has also led to the spectroscopic confirmation of 20 new
evolved massive stars selected based on their excess Paschen-$\alpha$
emission~\citep{mau10c}. One of these new discoveries is a rare
luminous blue variable (LBV) star, whose luminosity rivals that of the 
nearby ($\sim$ 7 ${\rm pc}$ in projection) Pistol star, one of the most massive stars
known in our Galaxy~\citep{mau10b}. 

In the present paper, we describe the data calibration and
analysis procedures, present the products, including a
catalog of detected sources, and construct mosaic maps, as well
as a list of Paschen-$\alpha$-emitting candidates.  These products
form a valuable database to study the massive star population in the GC.

The paper is organized as follows. We detail the procedures for
removing various instrumental effects in \S\ref{s:data}. In
\S\ref{s:anal}, we describe the source detection, the
construction of an extinction map, and the identifications of
Paschen-$\alpha$-emitting candidates, as well as the mosaics of the 1.9
$\mu$m, 1.87 $\mu$m, and  Paschen-$\alpha$ intensities. We
present our products in \S\ref{s:result} and discuss the nature
of the detected sources in \S\ref{s:dis}. We
summarize our results in \S\ref{s:summary}.

\section{Data Preparation}\label{s:data}
While details about the design of our survey can be found in
\citet{wan10}, Table~\ref{t:observation} summarizes its basic
parameters for ease of reference. Each orbit includes four pointing
positions, while each pointing position includes four dithered
exposures/images (in a square-wave dither pattern) for the two filters
(F187N and F190N), respectively. Therefore, 
 the entire survey consists of
4608 raw MULTIACCUM exposures 
(MAST data set IDs na131a-na136o) reduced initially to 
individual count-rate images. In addition, 
16 MULTIACCUM dark frames identical in sample
sequence and number of readouts to the GC exposures were 
obtained after the science exposures in each orbit during 
occultation, to provide contemporary, on-orbit calibration data 
and improve the instrumental background measurement. Our survey
represents the largest contiguous spatial scale mapping obtained with the
\hst/NICMOS camera. Although the survey has produced the highest 
resolution map of the intensity distribution at 1.87 $\mu$m and 1.90 $\mu$m
to date, our primary goal was to produce a photometrically accurate 
map of the Paschen-$\alpha$ 
emission throughout the survey region. In the 1\% filters used for this survey, accurate measurement of the 
Paschen-$\alpha$ emission requires that both the dominant bright 
stellar continuum and the instrumental background be carefully
removed. In this section, we describe the data
preparation required to achieve this goal. The data preparation is
based chiefly on the Image Reduction and Analysis Facility (IRAF).
In particular, the IRAF/STSDAS `calnica' and `Drizzle' packages, with
some procedural modifications, are used to produce images for each
individual position with either filter. We have also developed
our own routines in the Interactive Data Language (IDL) for the
background subtraction and for astrometric offset
correction.

\subsection{Calibration files update}\label{s:new}
We first apply the IRAF/STSDAS `calnica' to remove various instrumental
and cosmic-ray (CR)-induced  artifacts in the raw data of individual 
dithered exposures. Required inputs to the `calnica' program include 
dark, flat field and bad pixel reference files.  We do not use the 
STScI provided OPUS PIPELINE calibration files, but rather construct 
our own reference files based on either contemporaneous, or the 
most recently available, calibration data.

\begin{figure*}[!thb]
  \centerline{
       \epsfig{figure=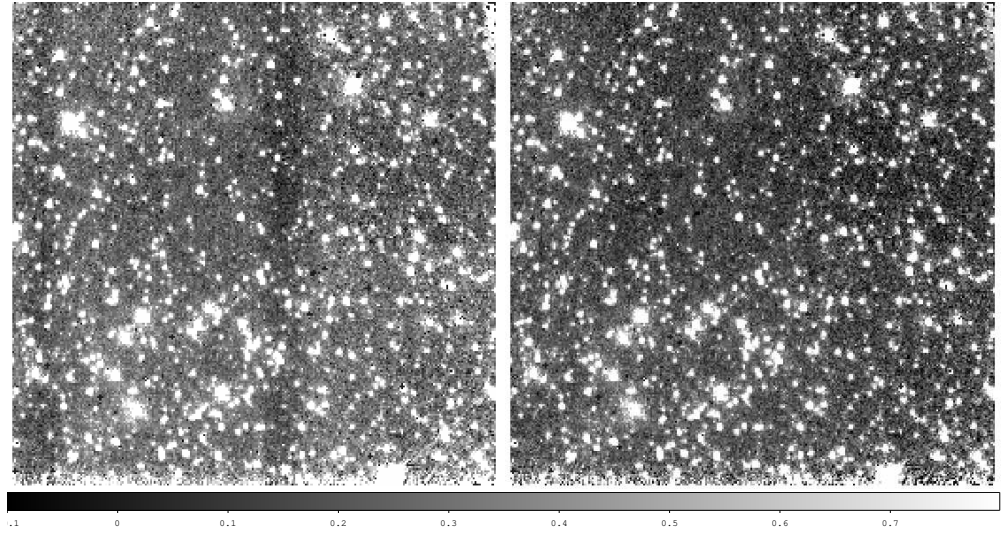,width=0.8\textwidth,angle=0}
    }
 \caption{Comparison of the outputs from `calnica' with the application of
the STScI provided dark file and the superdark produced with
 data obtained as part of our program (right panel). The left
 panel shows two obvious nearly vertical dark
          lanes.
          The fuzzy structures at the bottom rows of both images are due to the
          vignetting problem; these rows are removed from the subsequent
          data reduction.}
 \label{f:dark_compare}
 \end{figure*}

First, because the multiple components contributing to the 
instrumental dark signature can vary with the thermal state 
of the instrument and telescope on multi-orbit timescales, 
our dark frame acquisition strategy allowed for reference file 
acquisition and creation individually for each orbit.  The assemblage 
of dark data themselves, however, along with the NIC1 temperature
sensor data for a  proxy for the off-scale ${\sl NIC3}$ thermal sensor, 
indicates that the stability flanking all of our observations at a
level of $\le$ +/- 50
${\rm mK}$. We therefore choose to produce a single high S/N `superdark'
 by median combining our
dark exposures from all orbits. In order to check the reliability of
this `superdark', we examine the differences between output
files using this superdark and a dark file produced by only median-averaging the 16 dark
exposures in the same orbit. We select a single dithered exposure
(na133id1q) for this comparison, since it covers a sky location having low surface
brightness and dominated by the instrument background. 
While the small difference ($<5$\%) indicates the
similarity of these two dark files, the pixel value distribution of
the calibrated image with the `superdark' is much smoother than that
from the other dark file, suggesting the higher S/N of the
`superdark' file. The superdark significantly reduces the observed 
`Shading' effect --- a noiseless signal gradient in the
detector~\citep[NICMOS Data Handbook,][]{tha09}; the effect is readily apparent if the dark calibration file
provided by the
OPUS pipeline is used, which
was produced in 2002 (Fig.~\ref{f:dark_compare}).

Next, utilizing all of the 4068 dithered exposures obtained for the science survey, we
construct a new empirical mask file to identify the locations
of bad pixels (hot, cold or `grot') in ${\sl NIC3}$. We calculate the
intensity median and 68\% confidence error ($\sigma$) among the
pixels of each dithered exposure and record those pixels with
intensities deviating from the median by 1$\sigma$. We consider a
pixel to be bad if it is flagged in more than 75\% of the
exposures. In total, we identify 476 new bad pixels, in addition
to 695 in the origin mask file provided by STScI, which used a
more conservative threshold. We compare this new 
mask file with the science exposures and find that these new bad pixels
are real. These bad pixels together represent
0.7\% of the total number of ${\sl NIC3}$ pixels. The bad pixel mask also
includes the following: 1) the bottom 15 rows, which show a steep sensitivity
drop due to the vignetting problem~\citep[NICMOS Data Handbook,][]{tha09}, 2) a three-pixel-wide zone at the other three boundaries to avoid any potential edge
effects and 3) a 10$\times$50-pixel region in the
top-right corner, which was contaminated by the residual of the
amplifier readout and shows an unusual intensity enhancement, even
after we have subtracted the dark file
(Fig.~\ref{f:dark_compare}). All these bad pixels are removed from
the individual exposures before further processing.

The flat fields used in the initial pipeline processed data were taken 
in May 2002, during the SMOV3B operations, shortly after the installation
of the cyro-cooler. Since May 2002, flats for ${\sl NIC3}$ F187N and F190N were
obtained several times by the STScI. We retrieve ${\sl NIC3}$ F187N and F190N flat
field observations taken in September 2007 as part of the Institute's 
calibration program. These flat observations are the closest in time to our
program observations. We then process these observations for use as
the flat field images for our observations. By using the September 2007 
calibration data, we are able to correct for a persistent large-scale
flat field structure that is correlated over scale-lengths of ten or more 
pixels, and which lead to localized systematic errors in flat field 
corrections on the order of $\sim$ 2\%.  

\subsection{DC Offset Corrections}\label{s:pedestal}
In an individual exposure, obvious differences in background levels
exist among the four quadrants. This phenomenon is called the
`Pedestal' effect, a well-known \nicmos\ artifact due to
independent DC biases in the quadrants, each of which has its
own amplifier for data readout~\citep[NICMOS Data
Handbook,][]{tha09}. 
Within a quadrant, however, the
bias is constant. This effect must be
rectified to correctly map out the surface brightness distribution.

The IRAF routine `pedsub' is commonly used to correct for this
effect (before the application of the `Drizzle' package discussed
below). 
The routine
consists of three steps: 1) estimating the DC bias level
independently for each quadrant; 2) determining the DC bias
offsets among different quadrants by matching the surface
brightness smoothly across their boundaries; and 3) removing the
offsets from the respective quadrants. To do the matching in Step
2, one needs to choose (by trial and error) a method (median,
mean, or polynomial fitting). This approach works poorly, however,
when the intrinsic surface brightness changes abruptly across the
boundary between two adjacent quadrants (e.g., when the change
cannot be fitted well by any of the methods).

\begin{figure*}[!thb]
  \centerline{
       \epsfig{figure=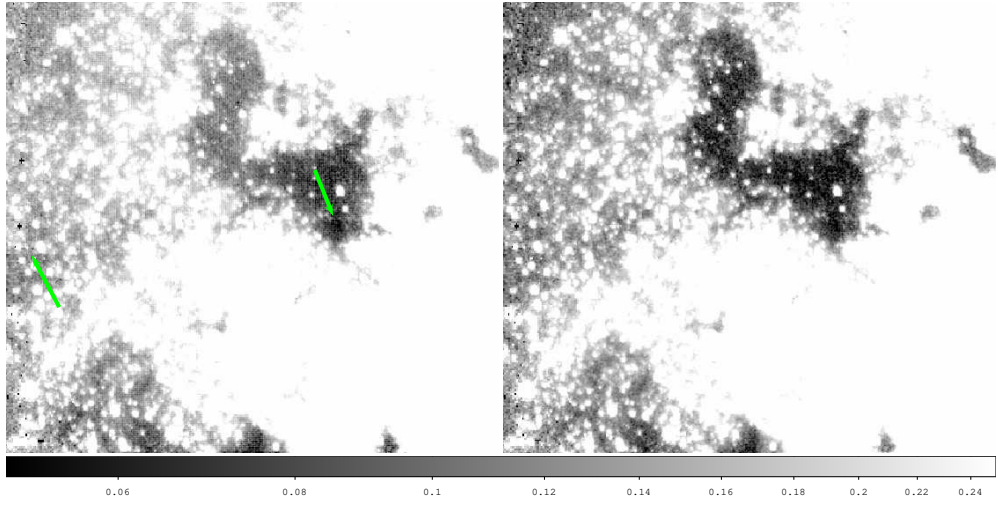,width=0.8\textwidth,angle=0}
    }
 \caption{Comparison between the position images that use either
the `Pedsub' step 2 (left panel) or our self-calibration method
(right panel) in two overlapping regions of quadrants. The green arrows in
the left panel point to
         the places that illustrate flux jumps at the boundaries of
         the quadrants of individual dithered exposures. This
Pedestal effect becomes much more prominent in an image containing net
Paschen-$\alpha$ emission.}
 \label{f:pedsub_compare}
 \end{figure*}

\begin{figure*}[!thb]
  \centerline{
       \epsfig{figure=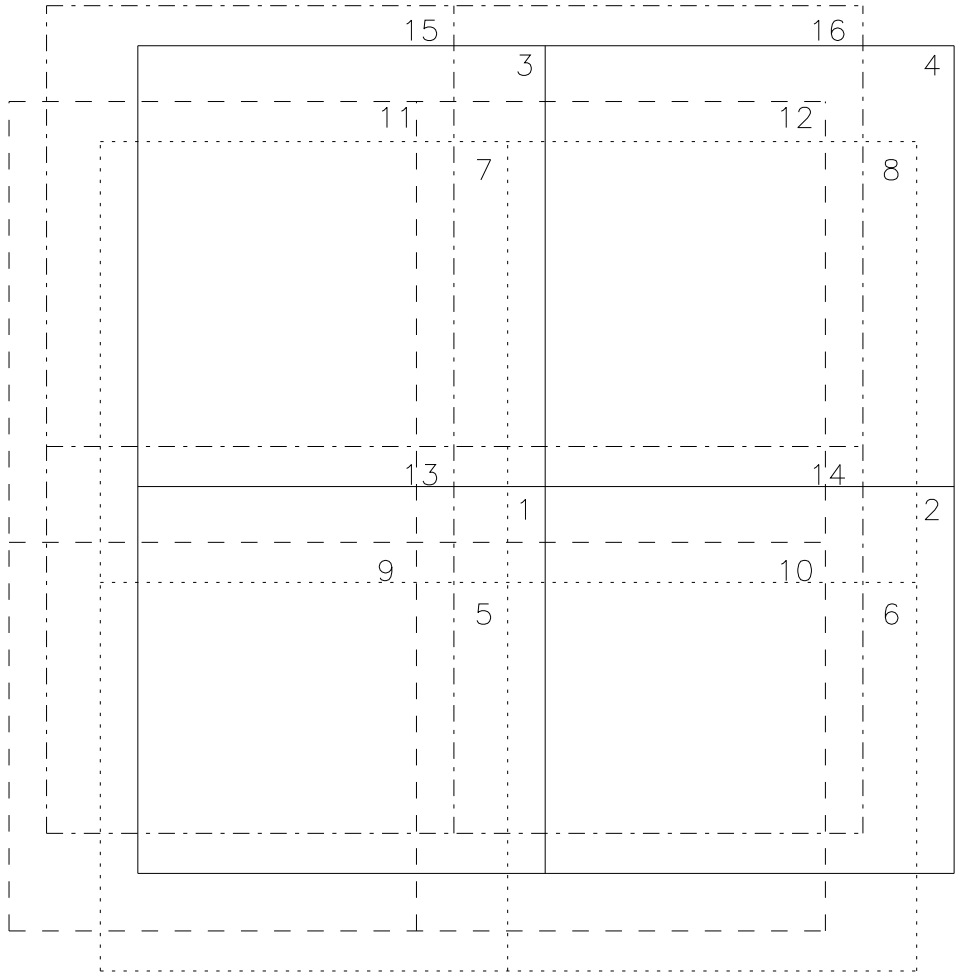,width=0.8\textwidth,angle=0}
       }
 \caption{Illustration of the four-point dithering pattern. For each position,
the dithering starts with the right-most exposure and proceeds
clockwise. The four exposures are outlined separately by solid,
dotted, dashed, and dotted-dashed boxes. Each exposure consists of
four quadrants. Each of the 16 quadrants (4 for each exposure) is
uniquely labelled for ease of discussion in \S\ref{s:pedestal}.}
 \label{f:drizzle}
 \end{figure*}

We have thus developed an effective self-calibration method to determine the DC
bias offsets from the overlapping regions among the four dithered exposures taken at each pointing position (diagrammed in Fig.~\ref{f:drizzle}). Although individual
quadrants in a single exposure have independent sky coverage, they
become connected with each other via the square-wave dither pattern.
After aligning the four dithered exposures, we identify all overlap
regions (see \S\ref{s:drizzle}). There are a total of
48 different overlapping region pairs in each of the four dithered exposures. The
mean intensity in an overlapping region can be expressed as
$f$=$f_{s}$+$f_{DC}$ where $f_s$ is the sky background, and $f_{DC}$
is the DC bias offset.   Since $f_s$ should be the same in identical 
quadrants observing the same piece of sky, the measured 
difference provides a relative measurement of the DC bias offset level.  
In fact, we can simultaneously determine
all the offsets in the 16 quadrants contained within a single
pointing position,  by conducting a $\chi^2$ fit to the
differences in the 48 overlapping region pairs (see Appendix). While
only 15 of the 16 offsets are independent, we add another
constraint that their sum equals zero (i.e., no net bias
changes averaged across a position image). Without the trial and
error, as would be needed in the replaced Step 2 in `pedsub', our
self-calibration method provides an efficient method for removing the
`Pedestal' effect (Fig.~\ref{f:pedsub_compare}). Upon completion of
this step, the relative DC offsets between the quadrants of a single
 exposure, and between the four dithered images of a single pointing position
have been effectively set to a common DC level.  

Next, we need to remove the DC offsets between the four pointing
positions, 
within a single orbit and between the 144 different orbits.  
We adopt the same global fitting method as used
for the quadrant offset problem above to estimate the relative DC offset
differences among pointing positions within an orbit, and between the 144 different orbits.
We first calculate the mean
difference between two adjacent pointing positions and the statistical
error based on their overlapping region. This fitting methodology 
therefore leads to
575 linear equations. Again we add the additional requirement that the
sum of the corrections in the 576 positions to be zero. The
resultant solutions for the 576 offset corrections are added
back into the original position images, thereby effectively
normalizing all of the images to a single DC offset level.

In order to calibrate our normalized survey images to the true
background count rate, we use a set of very dark
regions in the F187N and F190N
bands to establish the zero levels (Fig.~\ref{f:total_color}). 
These regions which are observed to be extended and dark, against
the bright near-IR background of the GC, must be caused by
the very strong extinction of foreground molecular clouds
(i.e. close to the solar neighborhood). We assume there is no 
detectable astrophysical flux above the instrumental sensitivity 
over these regions. Therefore, any residual measured counts in
 these regions are due to the global DC offset level that is a 
result of the background minimization methodology discussed above. 
To minimize the effect of the statistical uncertainties in our
estimation, we fit the histogram of the intensities obtained in
pixels of the dark regions with a Gaussian. The fitted Gaussian
central intensity value is adopted as the absolute background
and is subtracted from all images. 

To convert from the instrumental count rate to physical flux intensities, in the final 
of the background-subtracted images at 1.87 and 1.90 $\mu$m,
we apply the scale factors (PHOTFNU) of 4.32 and 4.05
$\times10^{-5}$ {\rm Jy} (ADU $s^{-1}$)$^{-1}$, which
are obtained from the \hst\ \nicmos\ Photometric webpage.
\footnote{http://www.stsci.edu/hst/nicmos/performance/photometry/postncs\_keywords.html}

\subsection{Position image construction}\label{s:drizzle}
After removing the DC offsets from the entire survey as discussed above, we merge the four individual dithered
exposures to form a combined image, which we call the `position
image'. 
We do not use the standard IRAF `calnicb' for this task. As
mentioned in the \hst\ \nicmos\ Data Handbook~\citep{tha09}, `calnicb' does not
allow for pixel subsampling and hence provides no improvement of the image
resolution from the use of the dithering. This routine also does
not correct for geometric distortion (the pixel size as projected on
the sky in the X
direction is about 1.005 times larger than that in the Y direction for
${\sl NIC3}$). Instead, we use the IRAF `Drizzle' package to merge the
exposures. This package is widely utilized for such a task.
Allowing for the dithering resolution enhancement, `Drizzle' uses
a variable-pixel linear reconstruction algorithm, which is thought
to provide a balance between `interlacing' and `shift-and-add'
methods. The behavior of this algorithm is determined by the
parameter `pixfrac', which defines the fraction of the input
pixel size to be used for the output one. We fix the parameter to
be 0.75. With the correction for geometric distortion, an
output image is uniformly sampled to 0.101$\arcsec$ square pixels, half of
the original pixel size of ${\sl NIC3}$ (i.e. `SCALE'=0.5 in `Drizzle').
Step-by-step, `Drizzle'
first uses `crossdriz' to produce `cross-correlation' images
between the different exposures, then `shiftfind' to determine the
relative spatial shifts between the exposures, and finally
`loop\_driz' to merge the exposures into a single position image. The
pixel size of the new images below is 0.101$\arcsec$ .

The `Drizzle' package provides additional tools to identify
the outliers which are not recognized by `calnica'. Toward the same
line-of-sight within the four dithered exposures, 
these tools select out the outlier pixels which cannot be explained by the Poisson
uncertainty of the detector's electrons. These pixels are removed from the further merge
 process in `loop\_driz'. However, these tools could potentially identify
 the core of bright sources in several dithered exposures as cosmic-ray, due
 to the undersampling problem of the ${\sl NIC3}$ and could make us
 underestimate the intensity of these sources (see also \S~\ref{s:pal_emit}).

\subsection{Astrometry correction}\label{s:astrometry}
Before combining all the position images to form a mosaic map for
each filter, we need to correct for different astrometrical
uncertainties.
 We first
account for the relative shift between the two filter images of
each position, due to the cumulative uncertainties introduced by the small angle maneuvers associated with the relatively  large offset dithers. We calculate the shift using 50 or more relatively
isolated bright sources, detected in both filters. The largest
shift is $\sim$ 0.025$\arcsec$, consistent with the expected \hst\
astrometric accuracy, $\sim$ 0.02$\arcsec$, for observations with at least
one guide star locked and within a single orbit (HST Drizzle Handbook). Such a shift,
though small, could still significantly affect an effective
continuum subtraction (especially in the vicinities of relatively
bright sources) required to produce a high-quality
Paschen-$\alpha$ image. We thus correct for the shift for each
position by regridding the F187N image to the corresponding F190N
frame using IRAF `geotran' with interpolation. But we neglect any
relative shift between different positions in the same orbit. Such
a shift should be similar to that between the filters and thus
much less than the image pixel size. We directly merge the
four pointing positions to form a mosaic image for each orbit, using the
coordinate information in their fits headers.

\begin{figure*}[!thb]
  \centerline{
       \epsfig{figure=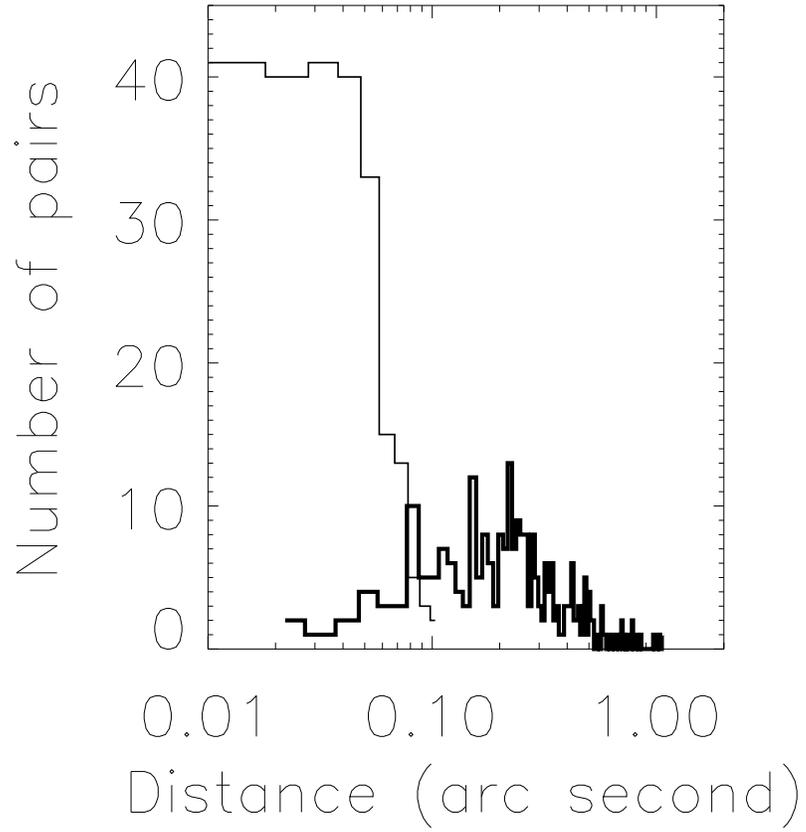,width=0.8\textwidth,angle=0}
       }
 \caption{Distributions of the relative spatial shifts between
adjacent orbit image pairs  before (thick line) and after (thin line)
the astrometric corrections.  }
 \label{f:dis}
 \end{figure*}

We need to correct for the relative shifts between orbit images. Each
of these images is created
by combining the four position images within each orbit with 
`Drizzle'. Because different orbits often used different
guide stars, we expect a relatively large astrometric uncertainty, which could be up to about 1$\arcsec$ and must be corrected for before creating a
final mosaic map of the entire survey region. According to the
`Drizzle handbook', the roll angle deviation of \hst\ is less than
0.003 degrees, i.e. 0.002$\arcsec$ at the edge of the ${\sl NIC3}$. Therefore,
we just include the $\alpha$ and $\delta$ shift, but no rotation deviation in our
astrometry correction process, for simplicity. We determine the relative
shifts among all 144 orbits in a way similar to that used for the
instrumental background DC
offsets (\S\ref{s:pedestal}). We estimate the relative shifts and
their errors (in both $\alpha$ and $\delta$ directions) between two adjacent
orbits via the cross-correlation in the overlapping region. A
total of 254 pairs of the relative shifts ($\Delta \alpha_{i,j}$ and
$\Delta \delta_{i,j}$) are thus obtained. The global fit is then
reduced to solve 2$\times$143=286 equations for the required $\alpha$
and $\delta$ corrections to be applied to the 143 orbit images.
Fig.~\ref{f:dis} compares the distribution of the relative spatial
shifts between adjacent orbit images ($\sqrt{\Delta
\alpha_{i,j}^2+\Delta \delta_{i,j}^2}$) before and after the correction. The
astrometric accuracy after the correction is sustantially improved;
the median and maximum of the shifts are 0.039$\arcsec$ and 0.107$\arcsec$,
representing the relative astrometric precision of our survey.

Finally, we calibrate our astrometrically aligned survey image to the
absolute astrometry of the Galactic center, by using the precisely
measured positions of SiO masers known within the central 1' around Sgr
A*~\citep{rei07}. These SiO masers, believed to arise from the
circumstellar envelopes of giant or supergiant stars, have
well-determined positions with uncertainties of only $\sim$ 1 mas
\citep{rei07}. We adopt the radio positions of the masers
determined in March 2006 and account for their proper motions
(Table 2 in \cite{rei07}). We identify 11 counterparts of our
F190N sources (see section~\ref{s:source}) among the 15 masers;
the remaining four (ID 11, 12, 14, 15; Table 1 of Reid at al. 2007)
are apparently below our source detection limit and are thus not
used. We then correct for the mean $\alpha$ and $\delta$ shifts estimated
between the radio and F190N positions of the masers 
[($\Delta\alpha$, $\Delta$$\delta$)=(0.03$\arcsec$,0.41$\arcsec$)]. After this
correction, the median and maximum spatial shifts between the
radio sources and their F190N counterparts are 0.03$\arcsec$ and 0.08$\arcsec$,
consistent with the residual astrometry uncertainty of our final
mosaic maps mentioned above. Therefore, the final median astrometric accuracy
of the mosaic is $\sim\sqrt{0.039^2+0.03^2}=0.049\arcsec$.

\section{Data Analysis}
\label{s:anal}

With the data cleaned, background-subtracted, and corrected for the astrometry,
we conduct data analysis to detect point-like sources, to construct
the extinction and Paschen-$\alpha$ maps, and to identify
Paschen-$\alpha$-emitting candidates.

\subsection{Source Detection}\label{s:source}

We use IDL program `Starfinder'
for the source detection~\citep{dio00}. This routine, based on point spread
function (PSF) fitting, is
well-suited to detect and extract the photometry of relatively faint
sources in a crowded field, as in the GC. Here we describe the key steps:

\subsubsection{PSF construction}\label{s:psf}
\begin{figure*}[!thb]
  \centerline{
       \epsfig{figure=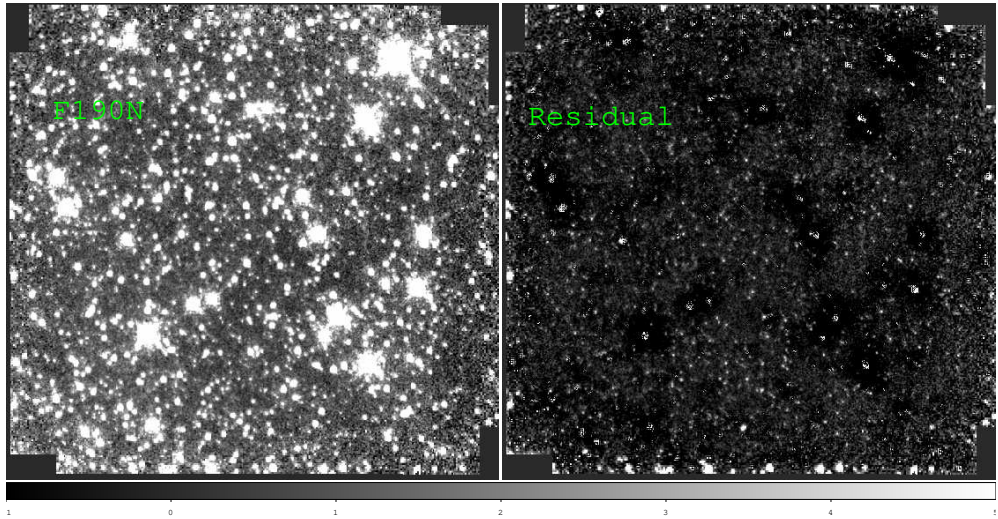,width=0.8\textwidth,angle=0}
       }
 \caption{Comparison of the original F190N image (left panel) and the
   residual image after removing the sources with our own PSF (right
   panel). We remove the sources near the edge from our catalog of
   each individual position because of the unreliable photometry. That
 is why these sources still exist in the residual image}
 \label{f:residual}
 \end{figure*}

Given the expected stalibity of the HST PSF, and the difficulty of
finding truly isolated PSF source stars in the crowded GC, we
construct a single PSF for use throughput the entire survey. We 
select 23 PSF template stars from our images
with $m_{K}<$ 7 from the 2MASS catalog which are isolated and have high quality
stellar counterparts in our survey. These stars are randomly
located in 22 position images (2 stars are located within the same image) 
and are relatively bright so that any
effect due to the presence of adjacent sources and/or small-scale
background fluctuation is small. For each star, an image of
128$\times$128 pixels ($\sim$ 12.8$\arcsec$$\times$12.8$\arcsec$) is
extracted from the resampled position image. The
resultant 23 star images are then normalized and median-averaged
(pixel-by-pixel) to form a 2-D PSF. We apply this PSF image to model and
subtract the contributions from any obvious adjacent sources in
each star image. The PSF image is then formed again. These last two
steps are iterated twice to minimize the effect of adjacent
sources on the final PSF image. In Fig.~\ref{f:residual}, we present the
original F190N image for one position and its residual image after
removing the sources to demonstrate the goodness of our PSF.  

\subsubsection{Local background noise level}\label{s:noise}
Another key input to the source detection routine is the average noise level
($\bar{\sigma}{_b}$) of the background in each position image.
To remove any large scale diffuse features,
we first subtract a median-filtered image (filter
size=1.2$\arcsec\sim$ 5
FWHM of the PSF; \S\ref{s:psf}) from the position image. We then fit
the histogram of residual intensity values with a Gaussian to
determine the pixel-to-pixel intensity dispersion, which we use
 as an estimate of $\bar{\sigma}{_b}$. This fit is not
sensitive to the high intensity tail that is due to the presence
of relatively bright stars. The median value of the dispersions is 0.017 ADU $s^{-1}$ $pixel^{-1}$,
consistent with the estimate from the \nicmos\ Expsoure Time 
Calculators (ETC), indicating
that they are mostly due to fluctuations in the instrumental
background. But positions near Sgr A* show exceptionally large
dispersions ($\sim$ 0.03-0.04 ADU $s^{-1}$ $pixel^{-1}$), because
of a high concentration of stars, resolved and unresolved.

\subsubsection{Source detection process}\label{s:detect}
With the PSF and $\bar\sigma_{b}$ as the inputs, we conduct the
source detection, which follows the following major steps: 1) Each
position image is median-filtered (filter size = 9$\times$FWHM of
the PSF) for a local background estimation; 2) A pixel brighter
than all 8 immediate surrounding ones and $\gtrsim
6\bar\sigma_{b}$ above the local background is identified to be a
source pixel candidate; 3) Such pixels are sorted into a
descending intensity order; 4) Starting with the highest intensity
one, the coefficient of the correlation between the PSF and the
surrounding sub-image that encloses the first diffraction ring 
is calculated; 5) A
source is declared if the correlation coefficient is
larger than 0.7; 6) This source with its centroid and photometry 
estimated from the PSF fitting is subtracted from the image before
considering the next pixel.
When one iteration is completed, all the detected sources are
subtracted from the original image and the same steps are repeated
again until no new source is detected.

All photometry measurements are performed on the position images prior
to construction of the full survey mosaic. For each individual
position image, we first detect the sources in the
images of F187N and F190N, independently. If two detections from
these two filter images are less than 0.1$\arcsec$ apart, we then consider that 
they are the same source. We also remove all detections that occur 
only in one filter. With the centroid fixed to the detections of the F190N, we rerun
`Starfinder' to obtain the photometry of the remaining sources in both
filters.

We remove or flag potentially problematic detections. Those detections 
with bad pixels within the first diffraction ring are flagged in 
Table.~\ref{t:source} (`Data Quality' = 1). We throw away
those detections that are less than 2$\times$FWHM away from an image
edge, which, if real, should mostly be detected with better
photometry in other overlapping position images. We also identify false detections
due to point-like peaks present in the PSF wings of bright
sources. For an arbitrary pair of adjacent detections (separation
$r <$2$\arcsec$), we estimate the PSF wing intensity of the brighter one
at the centroid position of the dim one. If this intensity is
greater than the net peak intensity of the dim
source, we mark it in Table.~\ref{t:source} (`Data Quality'= 2).

We merge the detections from all the position images in both filters
to form a single master source list. Two sources are considered to
be the same, if they are less than 0.25$\arcsec$ apart (FWHM of our PSF)
and are from different position images (after the astrometry is corrected;
\S\ref{s:astrometry}). The source parameters are adopted from the
detection that is farthest away from the position image edges.

\subsection{Source detection completeness limit}\label{s:det_limit}

\begin{figure*}[!thb]
  \centerline{
       \epsfig{figure=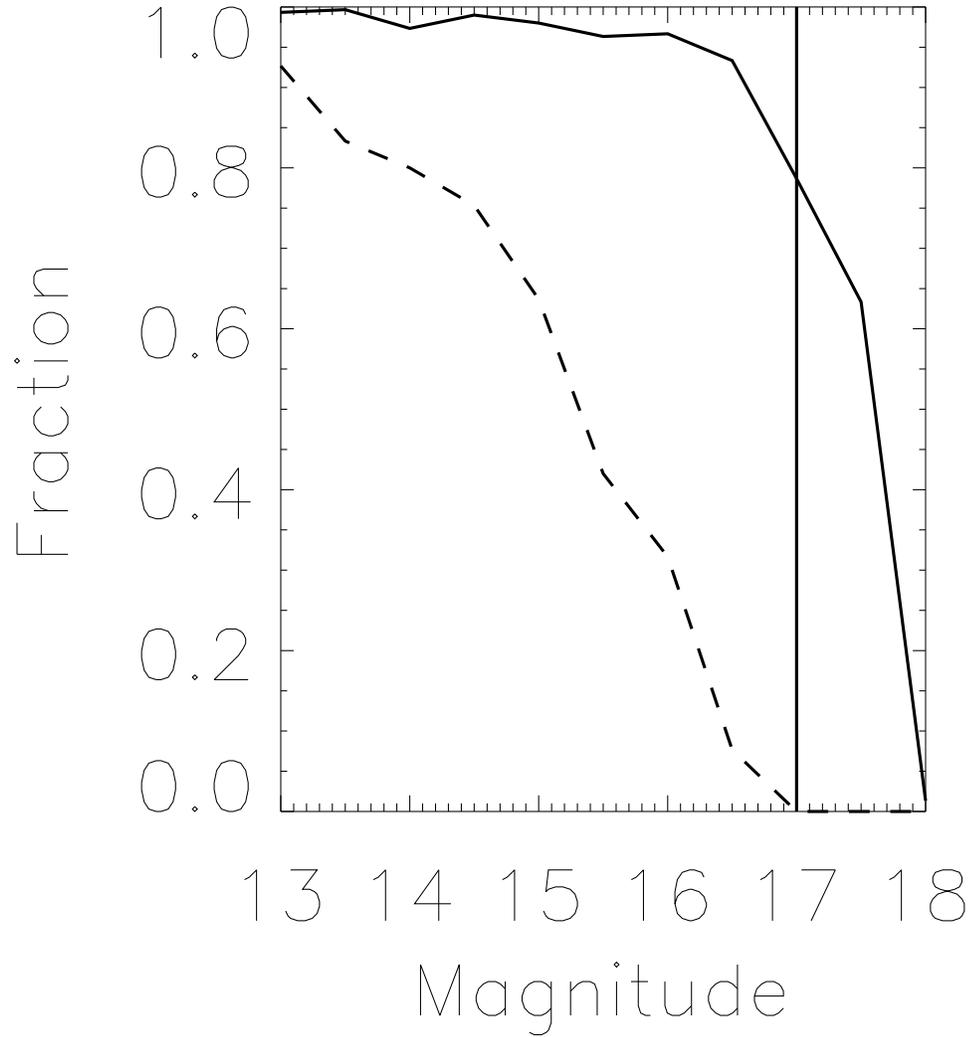,width=1.0\textwidth,angle=0}
       }
 \caption{The recovered fraction of the simulated sources as
a function of the F190N magnitude.
           The solid and dashed lines represent the positions with the
extremely low and high stellar densities (see \S\ref{s:det_limit}
for details). The vertical line at 17 magnitude represents the 50\%
incompleteness limit for most of the positions in our survey}
 \label{f:detec_limit}
 \end{figure*}

The source detection completeness varies from one position to
another, chiefly because of the variation in the stellar number
density. We estimate the average completeness in each position via
simulations. We simulate sources in the magnitude range from 13 to
18 with a bin size of 0.5, add them into each position image, and
rerun the source detection. Ten separate simulations, each
containing 30 sources, are conducted for each magnitude bin.
Fig.~\ref{f:detec_limit} illustrates the magnitude dependence of
the recovered fraction of simulated sources for the two
positions with the most extreme stellar number densities: the lowest
density position (GC-SURVEY-242) is on the Galactic north side
with a foreground dense molecular cloud, while the densest one
(GC-SURVEY-316) is near Sgr A*. For this latter position, the
fraction decreases quickly as the magnitude increases; the 50\%
incompleteness limit is about 15.5 mag. This limit increases to
17.5th mag for the lowest density case. For a more typical position in
our survey, the 50\% incompleteness limit is about 17th magnitude.

\subsection{Photometric Uncertainties}\label{s:error}
It is known that `Starfinder' severely underestimates actual
photometric uncertainty (Emiliano Diolaiti, private
communication). Therefore, we need to find a better way to accurately formulate 
and calculate the uncertainties based on our empirical data. The photometric
uncertainty should normally consist of at least two parts: 1) the Poisson
 fluctuation in the intensity ($\propto\sqrt{f}$, where f is the source
 intensity in units of counts) and 2) the local
background noise ($\propto\sigma_{b}$). In the case of ${\sl NIC3}$, we
also need to account for the `intrapixel' error, which is due to
the response variation over an individual detector pixel from the
center to the edge and due to the dead zones between the pixels.
This error depends on the degree of the undersampling of the PSF,
as is the case for individual ${\sl NIC3}$ exposures. Furthermore, when
converting the units ${\rm ADU/s}$ to ${\rm Jy}$, a systematic error ($f_{pe}$) is introduced
into the photometry, 0.0168 for F187N and 0.0051 for
F190N,\footnote{http://www.stsci.edu/hst/nicmos/performance/photometry/postncs\_keywords.html}
involving the use of the calibrated stars. Both of this systematic
error and the `intrapixel' error are reduced by a factor of
the square root of the number of the dithered exposures used in
constructing a position image. Putting all these together, we can express the total
model photometric uncertainty $\sigma_{f}$ (in units of ${\rm ADU/s}$) as
\begin{equation}\label{e:error}
\sigma_{f}^2-A*\sigma_{b}^2=\frac{f}{gain*t*N_{exp}}+\frac{(a*f)^2+(f_{pe}*f)^2}{N_{exp}},
\end{equation}
where $f$ (in units of ${\rm ADU/s}$) is the source flux, $t$ (${\rm s}$) is the
exposure time for individual dithered exposure, 
$N_{exp}$ is the total number of exposures, {\sl`gain'}
(6.5 ${\rm electron/ADU}$ for ${\sl NIC3}$) is an instrument parameter, and `A' is
the photometry extraction area in units of pixels. The local
background noise $\sigma_{b}$ (${\rm ADU/s/pixel}$) can be significantly
different from the $\bar{\sigma_{b}}$, the mean background over a
position image used in \S\ref{s:noise}, and can be quantified for
each detected source (see below). So only the coefficient $a$
needs to be calibrated for {\sl NIC3} .

\begin{figure*}[!thb]
  \centerline{
       \epsfig{figure=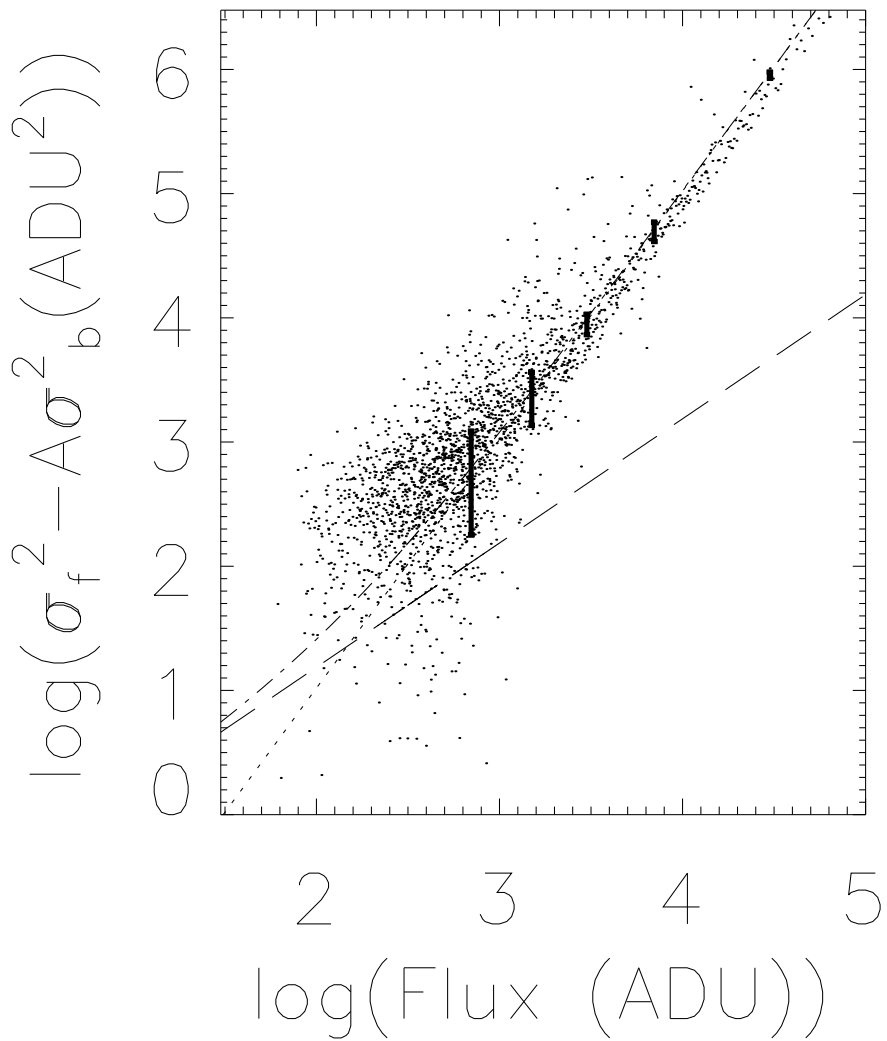,width=1.0\textwidth,angle=0}
       }
 \caption{Comparison between the empirically calculated and model
dispersions of the source fluxes. The term due to the local
background noise is estimated in individual sources and is
subtracted from the dispersions. The long dashed line and dotted
line represent the model contributions from the poisson
fluctuation and the intra-pixel sensitivity error, while the dash
dot line represents their total contribution. Representative error 
bars are illustrated at five flux levels.
          }
 \label{f:phot_err_2}
 \end{figure*}

We conduct this calibration by measuring the flux dispersions in
multiple dithered exposures of our detected sources. We select only those
sources that are not flagged and are detected in all four dithered
exposures of a position. We measure the source flux  in each
exposure, using the `daophot' package in IRAF (`Starfinder' is not suitable
for this purpose, because the PSF in a single exposure is
severely undersampled). The flux is extracted from an on-source
circle (radius=3 pixels or $\sim$ 0.6$\arcsec$, hence $A=3^2\pi$) after a
local background estimated in an annulus 1$\arcsec$-2$\arcsec$ (4-8 FWHM of the
PSF) is subtracted. The median of the standard dispersions within 
this background region from the four dithered exposures
gives $\sigma_{b}$. The mean and standard dispersion of the
fluxes ($f$ and $\sigma_{f}$) in the four dithered exposures are
then calculated. To conduct a $\chi^2$ fit of Eq.~\ref{e:error} to
the measurements, we further calculate the average and standard
dispersion of $\sigma_{f}^2-A*\sigma_{b}^2$ (the left side of
Eq.~\ref{e:error}) in each bin of $20$ sources obtained adaptively
from ranking their mean fluxes. The calculated values are presented in
Fig.~\ref{f:phot_err_2}. Since $\sigma_{f}$ is the dispersion of
the flux measurements among the dithered exposures, $N_{exp}=1$ and
$t=48$ {\rm s}. The best-fit ($\chi^2$/dof=4270./2624) then gives $a =
0.03$. Fig.~\ref{f:phot_err_2} compares the empirical measurements
and the best-fit model. The dispersion is dominated by the source
and background counting uncertainties in the low flux range and by
the `intrapixel' error in the high flux one.

With the calibrated $a$ value, we can now convert
Eq.~\ref{e:error} for estimating the photometric uncertainties of our
sources detected with `Starfinder' (\S\ref{s:detect}). In this
case, the error is for the mean source flux $f$ in a combined
position image, instead of the standard dispersion of the
detections in individual dithered exposures used in the calibration
above. The conversion can be done by setting (1) $N_{exp}$ equal
to the actual number of exposures covering each source (typically
four), (2) $A$ to $9^2$ ${\rm pixels}$, the region (including the first
diffraction ring) used to estimate the source flux in `Starfinder' above, and (3)
$\sigma_{b}$ to the standard dispersion in a box of  9$\times$9
FWHM in the `Starfinder' residual intensity images obtained after
excising all detected sources according to the PSF. The converted
Eq.~\ref{e:error} is then used to estimate the photometric uncertainty
for each of our detected sources.

A source flux ($f$) may be further expressed in magnitude as
\begin{equation}
m=-2.5 {\rm~log}(\frac{f}{f_{0}}),
\end{equation}
where we adopt the Vega zeropoint ($f_{0}$) as 803.8 {\rm Jy} (F187N) or 835.6 {\rm Jy} (F190N), as listed in the \nicmos\ photometric
keywords website.

We also need to have an absolute calibration of our photometry measurements
in the F187N and F190N bands. The measurements depend
on the goodness of the PSF as well as the calibration of the F187N and
F190N transmission curves. We conduct this  calibration by
comparing our flux measurements with predictions from stellar model 
spectral energy distribution (SED) fits to 2MASS JHK 
measurements, which have an excellent photometry
accuracy~\citep{skr06}. We first select
out 252 sources with J$<$14.1, H$<$11.6 and K$<$9.8, which insure that
these sources are 10\% brightest in all three 2MASS bands and 
show minimal flux confusion from other sources. We construct various SEDs  from
the ATLAS9 stellar atmosphere model~\citep{cas97}, together with
the surface temperature and gravity for stars of different types.\footnote{
http://www.stsci.edu/hst/observatory/etcs/etc\_user\_guide/1\_ref\_4\_ck.html}
For each source, we adopt the SED that gives the best $\chi^2$
fit to the JHK measurements with the extinction as a fitting
parameter (the extinction law of Nishiyama et al. 2009 is assumed). 26
among these 252 sources are chosen because of
their low extinction ($A_{K}<0.5$) and without substantial deviation ($<3\sigma$)
from the best-fit SES model predictions in F187N. The former
criterion is used to reduce the uncertainty in the extinction law that
is discuss in \S~\ref{s:dis}, while the latter 
one removes potential evolved massive
stars with significant emission in the 1.87 $\mu$m.
The fluxes from the best-fit SED are all
consistent with our measurements. The {\sl median} of the predicted to measured flux
ratios for the 26 sources is 1.016$\pm$0.009 (F187N) or
1.049$\pm$0.009 (F190N). This ratio is insensitive to the assumed 
extinction law ($<$1\%, which is used as a
measurement of the systematic uncertainty of the 
F187N to F190N flux ratio later) and is multiplied to the measured F187N or
F190N flux, respectively. If we do not modify the absolute photometry here, 
the flux ratio ($\frac{f_{1.87~{\rm \mu m}}}{f_{1.90~{\rm \mu m}}}$) would be
overestimated by 3\% and therefore the extinction derived in \S~\ref{s:extinction}
would be underestimated by 40\%.

\subsection{Ratio Map}\label{s:extinction}
To construct the net Paschen-$\alpha$ map, we need to
subtract the stellar continuum contribution in the
F187N band (\S\ref{s:palpha}). This contribution can be 
determined by using the observed intensity in the 
F190N images and the ratio of the F187N and 
F190N filters.  The two primary contributions to the variation in the F187N to
 F190N ratio are the interstellar extinction and the wavelength 
differences of the stellar continuum which varies slightly with
stellar type. In the GC, the extinction effect dominates, which can be estimated from the F187N to F190N flux ratios of our detected point sources.

Limited by the number statistics of the sources, we construct our
ratio map with a pixel size of 4$\arcsec$. For each pixel, we
obtain a median F187N to F190N flux ratio ($\overline{r}$) from 101
closest sources as the representative of the F187N to F190N
continuum flux ratio of the GC stellar light at that location. 
The number `101' is considered
to be a good balance between reducing the effect of the photometric
uncertainty in the median averaging and increasing the angular
resolution of our ratio map. As long as the bulk of the 101 sources
are low-mass stars located in the GC, the median averaging should not
be sensitive to a few outliers (e.g., massive stars or foreground
stars). The photometric uncertainty of the averaging ($\sigma_{\overline{r}}$) is $\frac{1}{\sqrt{101}}$ of the standard
dispersion estimated from the ranked values
(34 on each side of the median). While the median photometric error of
individual stars is $\sim 4\%$, the uncertainty after the average should then
be $\sim 0.4\%$.
We also record the maximum
angular distance ($d$) of these sources to the pixel center, which is 
as small as 4.3$\arcsec$ in the field close to
Sgr A*, where the surface density reaches $\sim$ 1.7
sources~{\rm arcsecond$^{-2}$}  and has a median of 9.2", averaged over
the whole survey area. This latter value may be
considered as the average resolution of the ratio map.

We use the ratio map to construct a high resolution extinction
map. Assuming 
an extinction law: A($\lambda$)=$\lambda^{-\Gamma}$, the
differential extinction at each pixel is
\begin{equation}\label{e:av}
A_{F187N}-A_{F190N}=((\frac{1.9003}{1.8748})^\Gamma-1)\times
A_{F190N}=-2.5\times {\rm log}[(\frac{f_{1.87~{\rm \mu m}}}{f_{1.90~\mu
    m}})_o/(\frac{f_{1.87~{\rm \mu m}}}{f_{1.90~{\rm \mu m}}})_m]
\end{equation}
where $(\frac{f_{1.87~{\rm \mu m}}}{f_{1.90~{\rm \mu m}}})_o$ and $(\frac{f_{1.87~{\rm \mu m}}}{f_{1.90~{\rm \mu m}}})_m$
are the observed and intrinsic (model) flux ratios, while
(1.9003/1.8748) is the ratio of the effective wavelengths of the two filters. As
discussed previously, 
$(\frac{f_{1.87~{\rm \mu m}}}{f_{1.90~{\rm \mu m}}})_m$= 1.015 for a
K0III  star and is not very sensitive to the exact stellar types assumed ($<$ 1\%
 for different types of evolved low mass stars). The above equation
 indicates that there is an anti-correlation between the  $\Gamma$ and
$A_{F190N}$.  To infer $A_{F190N}$ or equivalently $A_{K}$, we assume
$\Gamma = 2$~\citep{nis09}, which seems to be most consistent with the
RC magnitude location of stars in the GC (\S~\ref{s:dis}). If a
different extinction law is adoped, then the inferred extinction
is changed; e.g., $A_{F190N}$ or $A_{K}$ needs to be scaled
by a factor of 1.29/0.91/0.75 or 1.37/0.88/0.69 for
$\Gamma$=1.56/2.20/2.64, respectively~\citep{rie99,sch10,gos09}. 
 The uncertainty in the averaged photometry ($\sim 0.4\%$) also introduces
an  uncertainty in this extinction map: $\sim$0.22 mag or $\sim$0.16
mag in the $F190N$ or $K$ band.

We note that the above estimate
is problematic in regions with very strong foreground extinction.
Such regions can be strongly contaminated, if not dominated, by
foreground stars. Therefore, the derived ratio can be a poor
representation of stars in the GC. We identify such regions to
be those having a source number density ($\frac{101}{\pi d^2}$) $<$0.31
{\rm arcsecond$^{-2}$}, which corresponds to $\sim$ 2$\sigma$ below 0.38
arcsecond$^{-2}$, the density averaged across the survey field. At each
pixel of these regions, we adopt the extinction values
from~\citet{sch09} (using the conversion $A_{K}$=0.089$A_V$),
if it is greater than the one from our map. The extinction map of
\citet{sch09} is based on the {\sl Spitzer IRAC} photometry of red
giants and asymptotic giant branch stars and is sensitive to
strong extinction. But the map has a relative low resolution of $\sim~
2^\prime$, which is 12 times larger than the average resolution
of our extinction map, and it does not resolve compact dusty
clouds.

\begin{figure*}[!thb]
  \centerline{
       \epsfig{figure=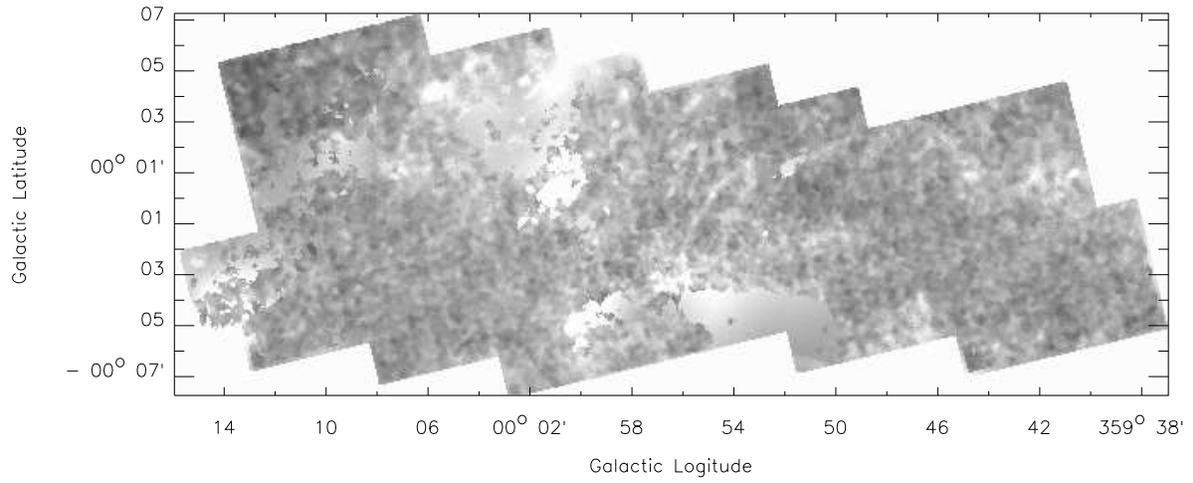,width=1\textwidth,angle=0}
       }
 \caption{Our adopted extinction map ($A_{F190N}$). White indicate the
   highest A$_{F190N}$ values. The minimum, maximum and median values of
   $A_{F190N}$ are 1.5, 6.1 and 3.0. The spatial resolution is $\sim~$9.2$\arcsec$}
 \label{f:av}
 \end{figure*}

Our adopted $A_{F190}$ extinction map (or the equivalent flux
ratio map) is presented in
Fig.~\ref{f:av} and is used for the construction of
Paschen-$\alpha$ images. One may infer $A_{K}$, using the conversion,
 $A_{K}=0.76\times A_{F190N}$, but is advised against a simple
 conversion to $A_V$ because the extinction law is very uncertain
 between the optical and infrared bands toward
the GC~\citep{rie99,nis09}. Because of the closeness of the two filters
in our survey, the readers should be aware of the large statistic and potentially systematic
uncertainty toward individual lines of sight of the extinction map. 
The criss-cross `textile' pattern in
Fig.~\ref{f:av} appears to be
artifacts of relatively large flux uncertainties in the overlapping regions
among the pointing positions. 

One may be concerned about the presence of stars with Paschen-$\alpha$
absorption lines at 1.87 $\mu$m, which may lead to an overestimation of the 
extinction. Such stars in a typical region are either too rare (e.g., O and B
stars) to affect our median average (used in constructing the
extinction map) or too faint (e.g., A-type Main-Sequence stars, which can have 
significant Paschen-$\alpha$ absorption) to be even detected 
individually. Only in the core of massive compact clusters may the crowded
 presence of such stars significantly affect the extinction estimate. We will examine this potential problem in a later paper. 

\subsection{Paschen-$\alpha$-emitting point sources}\label{s:pal_emit}
We identify a source to be a Paschen-$\alpha$-emitting candidate if its
flux ratio $r~(=\frac{f_{1.87~{\rm \mu m}}}{{f_{1.90~{\rm \mu m}}}})$ is significantly
above the local background value ($\overline{r}$). We define this excess as
\begin{equation}\label{e:pal}
r - \overline{r} >N_s \sigma_{tot},
\end{equation}
where
\begin{equation}
\sigma_{tot} = \sqrt{\sigma_{r}^2+\sigma_{\overline{r}}^2},
\end{equation}
and
\begin{equation}
\sigma_{r}^2=\frac{f_{1.87~{\rm \mu m}}}{f_{1.90~\mu
m}}\times\Big[\big(\frac{\sigma_f(1.87~{\rm \mu m})}{f_{1.87~\mu
m}})^2+\big(\frac{\sigma_f(1.90~{\rm \mu m})}{f_{1.90~\mu
m}}\big)^2\Big].
\end{equation}
We choose two values for the significance factor, $N_s$, of the 
excess: 4.5 and 3.5, resulting in the identifications of 197 and 341
potential Paschen-$\alpha$-emitting point sources among all
5.5$\times10^5$ stellar detections having cross-correlation values in both filters
larger than 0.8. We use a higher cross-correlation 
limit here (0.7 in \S~\ref{s:detect} ) to remove
sources with relative low detection quality. Statistically, the expected number of
spurious identifications among all the detected sources is about
2 and 83 for $N_s = 4.5$ and 3.5, respectively,
over the entire survey field. The latter (less conservative)
choice $of N_s$ is particularly useful for identifying
Paschen-$\alpha$-emitting candidates in small targeted regions (e.g.,
the known clusters), for which the expected number of spurious
sources would be negligible.

We individually examine these initial identifications in the F187N
and F190N images to flag potential systematic problems. First, with
the choice of $N_s$=4.5 (or 3.5), we
label 30 (42)  of the identified Paschen-$\alpha$-emitting candidates 
with `Problem Index'=1 in 
Table.~\ref{t:pal_sou_second}, since their total
fluxes within the central 5$\times$5 pixel in the calibrated 
F190N images are reduced by 1\%, 
compared with the images produced without
`cosmic-ray removal' steps in the `Drizzle' package (see
\S\ref{s:drizzle}). 
Second, each of additional 5 (10) sources has a
neighbor within 0.3$\arcsec$ (i.e. $\sim$ 1 FWHM of our PSF) and with  a
comparable flux (within a factor of 10). The photometric accuracy of
these candidates is somewhat problematic. Therefore, they are labeled 
with `Problem Index'=2 in 
Table.~\ref{t:pal_sou_second}. Third, our visual examination removes another 10
sources, the photometry of which are likely affected by the 
nearby bright sources (`Problem Index'=3 
in Table.~\ref{t:pal_sou_second}).

\subsection{Paschen-$\alpha$ image}\label{s:palpha}
\begin{figure*}[!thb]
  \centerline{
       \epsfig{figure=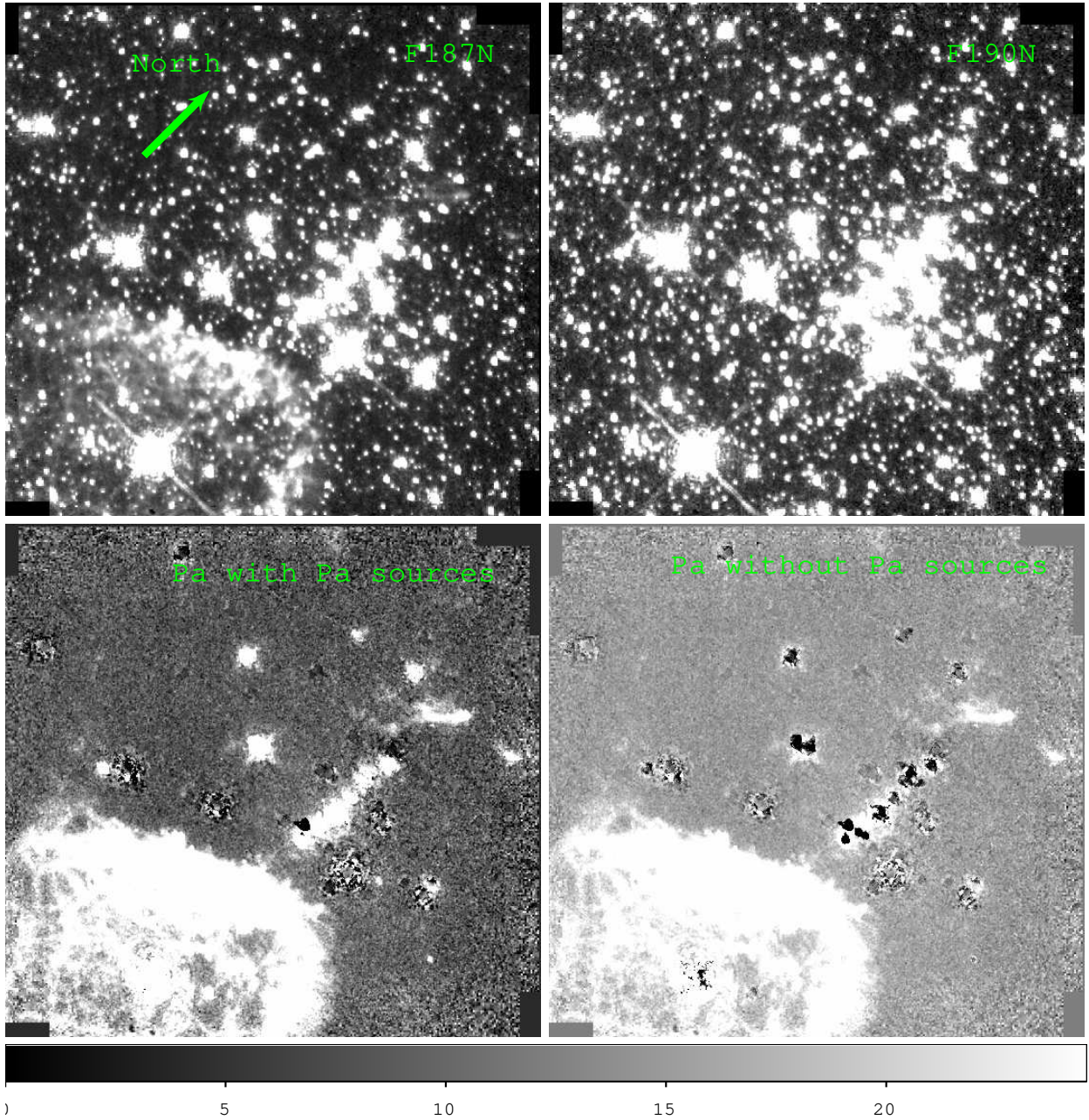,width=1\textwidth,angle=0}
       }
 \caption{The F187N (upper left) and F190N (upper right) images and
   differenced images with and without
   Paschen-$\alpha$ emitting sources (lower left and lower right, see \S\ref{s:pal_emit})
          of the position `GC-SURVEY-72' (in detector coordinates), which includes the central part of the
          Quintuplet cluster, the Pistol star and the Pistol Nebula. 
         Diffuse Paschen-$\alpha$ emission from the Pistol Nebula dominates the Paschen-$\alpha$ images.}
 \label{f:palpha_indi}
 \end{figure*}

\begin{figure*}[!thb]
  \centerline{
       \epsfig{figure=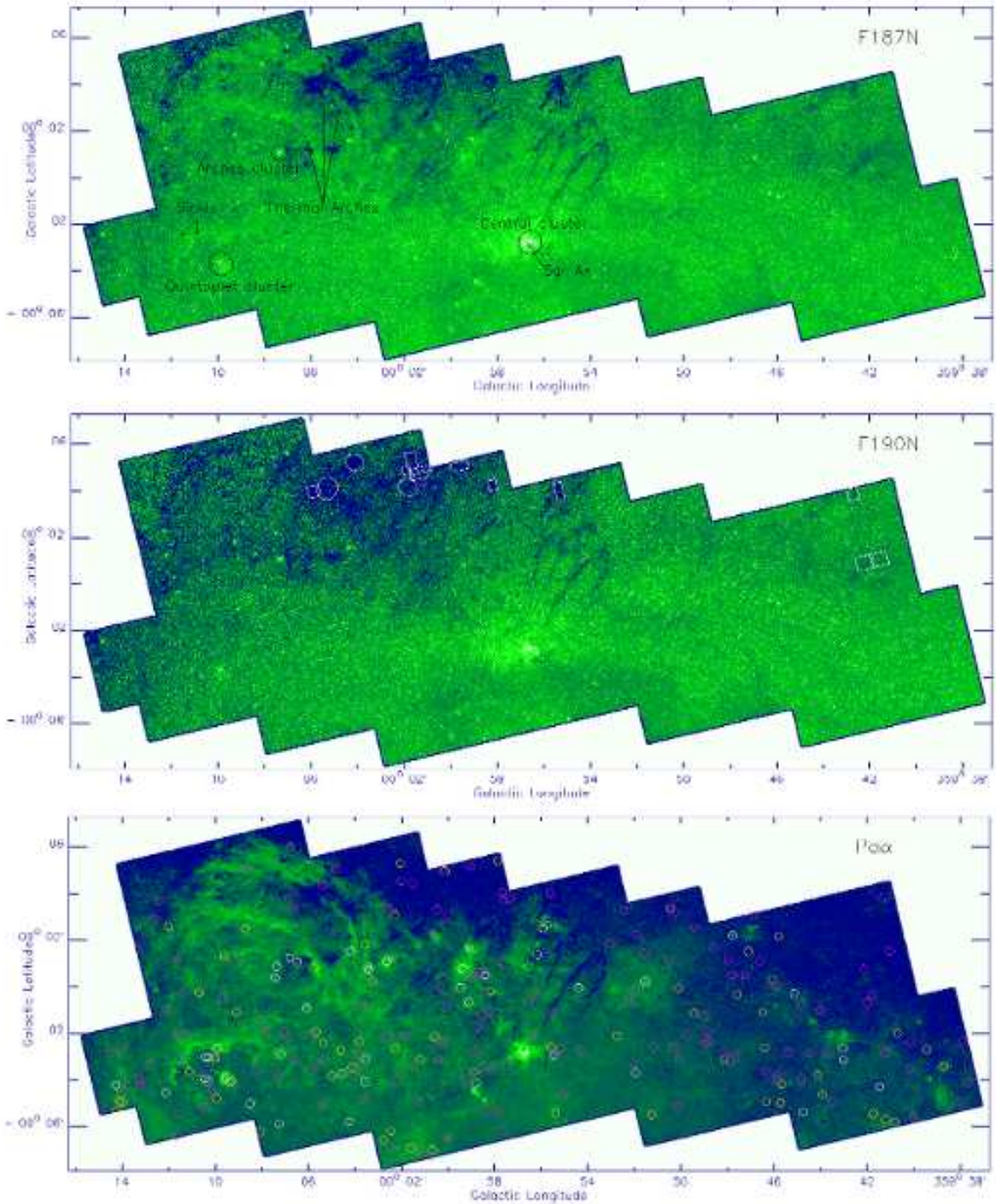,width=1\textwidth,angle=0}
       }
 \caption{Mosaic images of the F187N, F190N and
Paschen-$\alpha$ intensities. In the upper panel, several
well-known objects in the GC are marked. White circles and boxes
in the middle panel enclose regions used to calculate the absolute
background level in the images (see \S\ref{s:pedestal}). The white,
yellow and pink circles
in the lower panel represent the spectroscopically confirmed, unconfirmed primary and secondary Paschen-$\alpha$ emitting 
sources, respectively (see \S\ref{s:pal_emit}). For clarity, the
sources within the cores of the three clusters have not been overlaid.(You can find the full resolution image in www.astro.umass.edu/~dongh/paper/HST-GC-SURVEY/fig10.ps)}
 \label{f:total_color}
 \end{figure*}

The final Paschen-$\alpha$ intensity map is created by removing the
stellar continuum from the F187N image, using the F190N image and the
F187N/F190N point source ratios. The stellar contribution is derived from
the F190N intensity image by multiplying a corresponding scale
image, which depends on the local intrinsic stellar spectral shape
and the line-of-sight extinction determined as described in
\S\ref{s:extinction}.
 Because of the closeness of the wavelength of the two 
narrow bands, this dependence is
generally weak ($\lesssim 10\%$ for the expected extinction range
 over the survey area and much smaller for various
stellar types). Nevertheless, to map out low surface brightness
diffuse Paschen-$\alpha$ emission, we need to account for the
dependence, especially for pixels that are  affected by relatively
bright sources. We construct two kinds of Paschen-$\alpha$ images,
with or without the Paschen-$\alpha$ sources  ($N_s > 3$)
retained. We denote the latter as denote
the diffuse Paschen-$\alpha$ image. We adopt the `Spatially
Variable Scale' method used by \citet{sco03} to calculate the
scale factor for each individual source that needs to be excised.
This method involves many steps, including the allocation of
affected pixels to a source and the calculation of its
F187N-to-F190N flux ratio. The scale factor for a `field' pixel,
which is not significantly affected by sources, is directly
inferred from the ratio map constructed in \S\ref{s:extinction}.
This approach adaptively accounts for the spatial
extinction variation, which slightly differs from that used in \citet{sco03},
where an average flux ratio of detected sources is applied to an
entire position image. The product of the constructed scale map and the
F190N image is then subtracted from the corresponding F187N image
to map out the Paschen-$\alpha$ intensity. Finally, small-scale
residuals which deviate from the local median values by more
than 50\% error (due mainly to photon counting fluctuations around
removed sources) are replaced by the interpolation across neighboring
pixels in the resultant image, as was done in \citet{sco03}. As a
close-up demonstration, Fig~\ref{f:palpha_indi} shows the F187N, F190N
and Paschen-$\alpha$ images with and without the Paschen-$\alpha$
emitting sources in the `GC-SURVEY-72' position, which contains
the Quintuplet cluster, as well as the Pistol star and its Nebula.

\section{Products}\label{s:result}
Fig.~\ref{f:total_color} illustrates the products of the above
data analysis, including the 1.87 $\mu$m, 1.90 $\mu$m, and
Paschen-$\alpha$ mosaic maps constructed for the entire survey
field. The diffuse Paschen-$\alpha$ map has been presented in
\citet{wan10}. 

\begin{figure*}[!thb]
  \centerline{
       \epsfig{figure=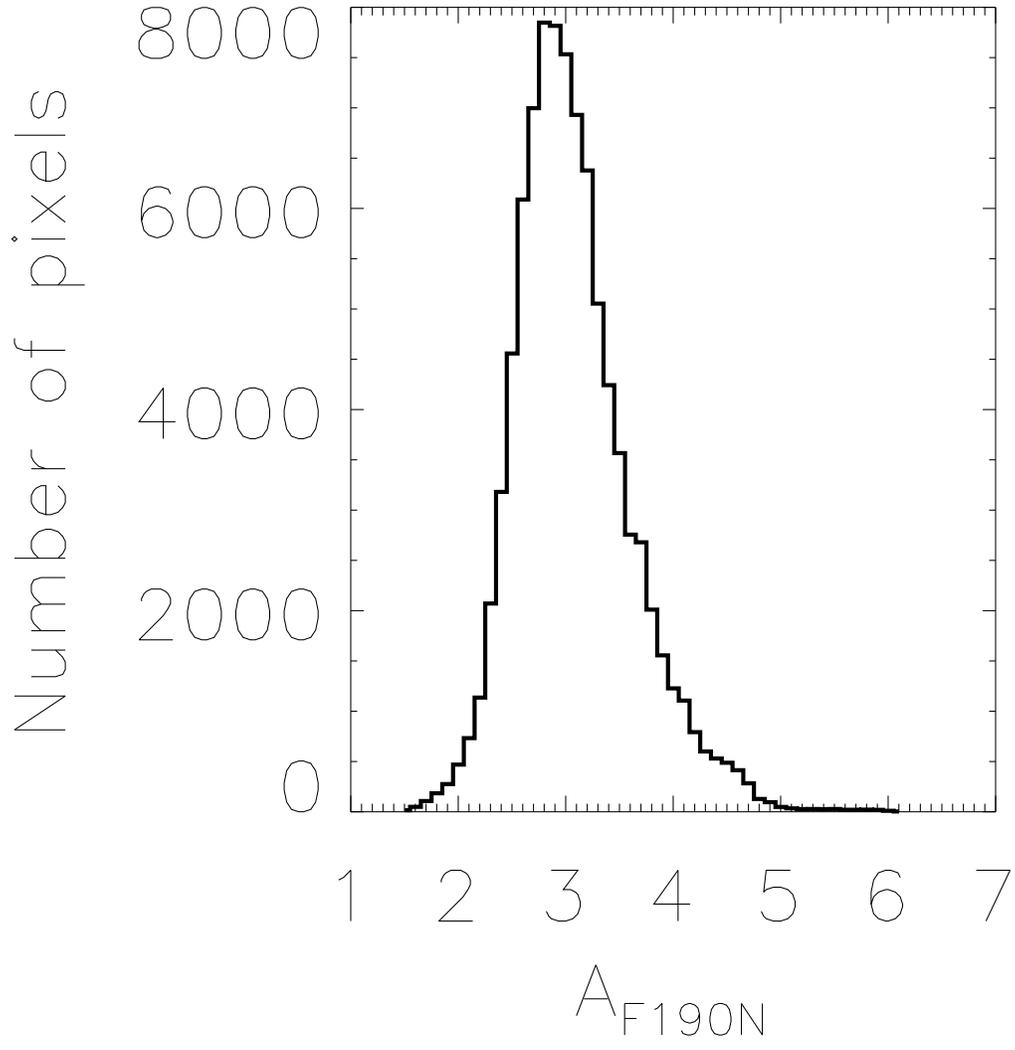,width=1\textwidth,angle=0}
       }
 \caption{A$_{F190N}$ distribution of the adopted extinction map.}
 \label{f:av_dis}
 \end{figure*}

Fig.~\ref{f:av} shows our adopted composite extinction
map, which represents the first large-scale subarcmin-resolution
measurement for much of the survey area.
Fig.~\ref{f:av_dis} further presents the $A_{F190N}$ histogram
constructed from the map. The median value ($A_{F190N}$) is 3.05, while the peak
in Fig.~\ref{f:av_dis} represents $A_{F190N}$=2.92$\pm$0.01,
corresponding to $A_{K}=2.22\pm0.01$ (using the extinction law of
Nishiyama et al 2009). If we adopt the extinction law of~\citet{rie99},
$A_{K}$ increases to 3.03$\pm$0.01, which is consistent with 
the average extinction ($A_{K}$=3.28$\pm$0.45) 
derived from stellar observations in 15 different regions within
the \gc\ by~\citet{cot00}. Our estimated extinctions toward the Arches,
Quintuplet and Central clusters are also consistent with other
independent observations to within 10\%~\citep{sto02,fig99a,sco03}, if
we adopt the slopes of the extinction laws they used.

In total, we detect 570,532 point-like sources in both the F187N and F190N bands
above a threshold of 6$\sigma$. Table~\ref{t:source} presents the
parameters for a sample of these sources, while the complete
catalog is published online only. Among the sources, 9,662
are flagged because of their proximity to relatively bright
sources or because of nearby bad pixels 
(see \S\ref{s:detect} and note to Table~\ref{t:source}).
 
Table~\ref{t:pal_sou} presents the 152 sources that are the most reliable
Paschen-$\alpha$-emitting candidates, which are identified with
$S/N > 4.5$ (see \S~\ref{s:pal_emit}) and with no flag for potential
systematic problems. In addition, we list tentative candidates in
Table~\ref{t:pal_sou_second}: those with problem flags and S/N $>$
4.5, and those having 3 $<$ S/N $<$ 4.5.

\section{Discussion}\label{s:dis}
\begin{figure*}[!thb]
  \centerline{
      \epsfig{figure=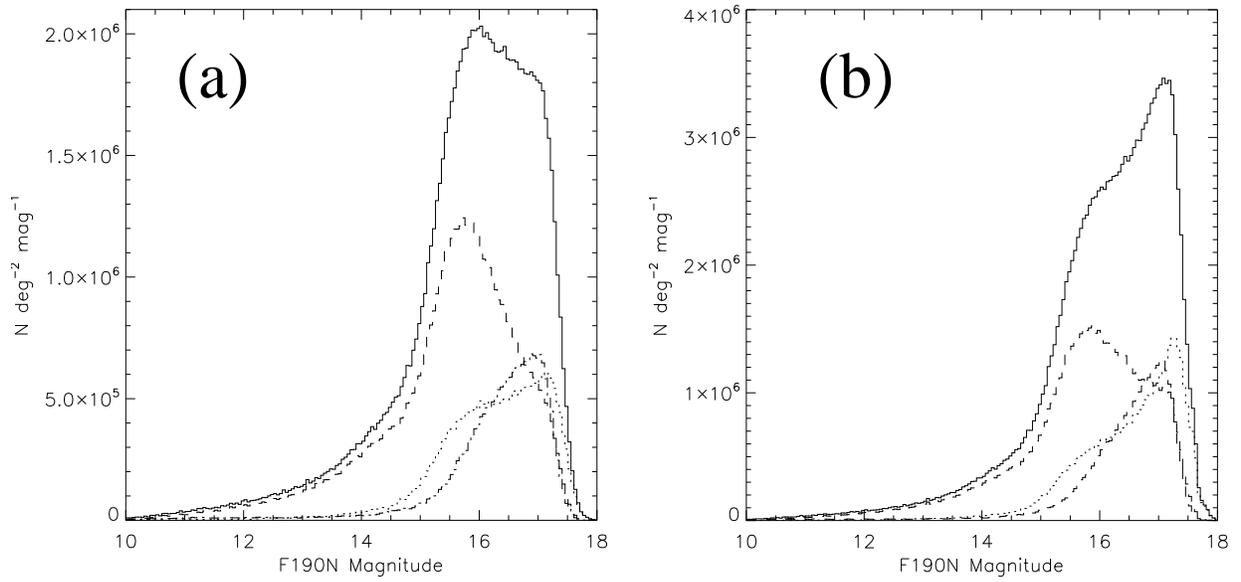,width=1.0\textwidth,angle=0}
       }
 \caption{1.90 $\mu$m magnitude distributions : (a) all sources
 (solid line) and only those in the color ranges $A_{K}<2$ (dotted),
$2<A_{K}<$5 (dashed), and $A_{K} > 5$ (dash dot); (b) The same as
(a), but approximately corrected for the detection limit. }
 \label{f:tot_lum_vsHK}
 \end{figure*}

The full data set presented in this paper contains a wealth of
information on the diffuse ionized gas, the stellar population, and
the extinction distribution toward the GC. Here, 
we focus on the statistical properties of the stellar population; the
data alone is typically insufficient for the study of individual stars. 
Fig.~\ref{f:tot_lum_vsHK} shows the 1.90 $\mu$m magnitude
distribution of our detected sources. We
 further roughly group these sources into the `foreground'
($A_{F190N}<1.8$ inferred from their {\sl individual} F187N to F190N flux
ratios) and `background' ($A_{F190N}$$>4.7$) as well as `GC' components
(1.8$<$$A_{F190N}$$<4.7$). The extinction range adopted for the GC
component approximately corresponds to 20$<$$A_{V}$$<$50, which is the
same as that estimated for 
the ionized gas within Sgr A West by~\citet{sco03}, 
who adopted the extinction law of~\citet{rie99}. 
The total source numbers are 1.4$\times10^5$, 1.2$\times10^5$ and 
3.1$\times10^5$ in
the `foreground', `background', and `GC' components,
respectively. This grouping is not meant to be precise,
particularly in the consideration of the closeness of the two narrow bands used to infer
the extinction and the uncertainty in the photometry.
Nevertheless, the components show distinct characteristics in
their magnitude distributions, as shown in
Fig.~\ref{f:tot_lum_vsHK}. The magnitude distributions of the
`foreground' and `background' components peak at $\sim~17$th
mag, which is mostly due to the decreasing detection fraction
toward the fainter end. In contrast, the distribution of the
`GC' component peaks at 15.80$\pm$0.01, which is far too bright  to be
due to the source detection limit variation (as demonstrated in
Fig.~\ref{f:tot_lum_vsHK}b). We fit a gaussian distribution to
this peak and obtain a width of 0.67 magnitude,
 which cannot be explained by the photometric
uncertainties of the sources within this magnitude range ($\delta
m_{F190N}\sim~0.04$). For a prominant old stellar population ($>$2
Gyr) as expected in the GC, the most probable
explanation for the peak is the presence of the RC stars, which represent a concentration in the color magnitude diagram~\citep{gro02}.   

Here we check how this RC  explanation is consistent with 
the peak of the 1.9 $\mu$m magnitude distribution of the
 GC stars, depending on the specific extinction law assumed (see also 
\S~\ref{s:extinction}). The Padova stellar evolutionary
tracks show that the RC peak is located at $M_{F190N}=-1.55$ for a 
2 Gyr old stellar population with the solar
metallicity. We adopted the
same distance modulus, 14.52$\pm$0.04, as useed by Sch{\"o}del et
al. (2010). The typical extinction toward
the GC in F190N is 2.92$\pm$0.01, as obtained in \S~\ref{s:result}, 
assuming the extinction law of~\citet{nis09}. Fig. 13 compares the
model and observed magnitude distributions. The RC peak locations
of the model (15.89$\pm$0.44) and observations are consistent with
each other. In contrast, the RC peak locations predicted from assuming other
 extinction laws seem to less consistent with the observed one
 (16.74$\pm$0.51, Rieke 1999; 15.63$\pm$0.41, Sch{\"o}del et al. 2010;
 15.16$\pm$0.35, Gosling et al. 2009).  The
uncertainties in the peak location of these models 
are derived from the Eqn. 3 by using 1\% systematic
error of the F187N to F190N flux ratio (see
\S~\ref{s:error}). Therefore, we can see that the Nishiyama's
extinction law best
matches our data.

Fig.~\ref{f:det_limit} further compares  the GC F190N magnitude
contours with stellar evolution tracks,
which are obtained from Girardi et al. (2000) for masses in the range of
0.15-7 $M_{\odot}$  and
from Bressan et al.(1993) and Fagotto et al.(1994a, 1994b) for 9-120
$M_{\odot}$. In the calculation of the F190N magnitude contours,
we have used the line-blanketed stellar atmosphere spectra from ATLAS 9
model~\citep[][ and references therein]{cas97}. It is clear 
that the majority of the GC sources with the limiting magnitudes as
discussed in \S~\ref{s:det_limit}
should be mostly evolved
low-mass stars, although a significant population can be Main-Sequence
(MS) stars with
masses $\gtrsim$ 5 (or typically 7) $M_{\odot}$ (i.e. stellar type
earlier than B5 or B3).

The `background' and `foreground' components mostly
represent the integrated stellar populations along the line of sight 
in the field. With even
larger extinctions and distances than the GC stars,
`background' stars should also be mostly evolved
(hence intrinsically bright) stars. In comparison,
the `foreground' component is
likely a mixture of MS and more evolved stars. In particular,
the `foreground' distribution shows a knee structure between 15th and 16th mag,
which is on the fainter side of the red clump peak. The fainter
stars in this structure
should mostly be MS and/or subgiants in the foreground Galactic disk.

Our Paschen-$\alpha$ candidate catalog is also contaminated by
foreground sources. It is difficult to estimate this contribution
based on our data alone. The follow-up spectroscopic observations
(\citealt{mau10c}) have shown that two of the 20 confirmed emission
line sources appear to be the foreground of the GC. These two
sources (stellar type O4-6I and B0I-2I) all have low extinction
($A_{K}\sim~1$) and show several He I (2.122$\mu$m, 4S-3P) and H I
(2.166$\mu$m, Br$\gamma$) lines; An apparent 1.87$\mu$m intensity excess could
then be due to the He I 4F-3D transitions or to the Paschen-$\alpha$ line. Our catalog is further
contaminated by non-emission-line foreground stars. For such a
star with a smaller line-of-sight extinction  than what is assumed 
for GC stars, the Paschen-$\alpha$
emission excess can be slightly overestimated. In the extreme case
of no extinction, $r\sim$1.015, instead of 0.942 for $A_{Ks}
\sim~2.22$ (Eq.~\ref{e:av}). This small overestimation can lead to
additional spurious identifications: about 2 and 83 for $N_s =
4.5$ and 3.5, accounting for the photometric uncertainties of the
sources with $r>1$, i.e. $A_{Ks}<0.44$.

To further investigate the line-of-sight locations of our Paschen-$\alpha$
emission sources,  we also use 2MASS and SIRIUS catalogs~\citep{skr06,nis06}
to identify 
foreground stars which are assumed to have $m_{H}-m_{K}<1$ or
$A_{Ks}<1.5$. We first search for the counterparts of our 
Paschen-$\alpha$ emitting sources from the SIRIUS catalog with both $m_H$ 
and $m_K$ measurements. The 2MASS catalog is used as a supplementary. The stars
within the three large clusters do not have reliable photometry in the
2MASS and SIRIUS catalogs due to low angular resolutions of these surveys; however,
the clusters have been studied in depth
elsewhere~\citep[e.g.][]{fig99a,fig02,pau06}. 
In the field regions, we find that 2 and 8 Paschen-$\alpha$ candidates
are likely foreground stars (H-K $<$ 1) based on the 2MASS and SIRIUS catalogs,
respectively. 
The two 2MASS sources
have been studied in~\citet{mau10c}, as discussed above, and are in fact
emission line stars. 
 Therefore, although the eight stars with
counterparts in the SIRIUS catalog are likely to be in the foreground  of the GC, we can not
exclude the possibility that they are still evolved massive stars in the Galactic 
disk. Further spectroscopic observations are 
needed to identify their origin. 
   
Our survey identifies almost all of the \gc\ massive stars with strong
line emission found in 
previous studies.
We include in Table~\ref{t:pal_sou} spectroscopically identified 
counterparts of our Paschen-$\alpha$ emitting sources, both in the three 
clusters and in the field regions. In Table~\ref{t:comparison}, we 
compare our detections with the stars that have been 
identified spectroscopically in individual clusters or nearby. 
We assume that massive stars within 3$r_c$ (see Table.~\ref{t:comparison})
 are cluster members. Our Paschen-$\alpha$ detections
(Table~\ref{t:comparison}) recover all 14 sources having the largest equivalent 
widths of Pachen-$\alpha$ line in~\citet{fig02}, either WNL (WN7-9) or OIf$^{+}$~\citep{mar08}. The other massive stars, which are still on or 
just leaving the MS, tend to show featureless spectra or even 
absorption lines and are not detected as Paschen-$\alpha$ emitting sources.
In the Quintuplet, we missed six of 19 WR stars identified 
in~\citet{lie09}, five of which 
are the Quintuplet-proper members~\citep{fig99a}, which all 
lack spectroscopic features in the K band. The sixth star has a similar 
spectrum. For the two LBV stars appearing in previous
literatures~\citep[such as, ][]{fig99a}, we found the Pistol star, but not
qF362~\citep[see ][ for more discussion on this unusual star]{mau10b}.

In the Sgr A West region, source confusion and a significant unresolved background stellar component severely limited our detection of Paschen-$\alpha$ 
emitting sources. Only 
17 of known 31 WR and OIf$^{+}$ stars in the Central Cluster are detected in our 
survey with $N_s>3.5$. 
Because the IRS 13 E complex is not well resolved in our survey, the WR 
stars, IRS 13 E4 and IRS 13E2~\citep[E48 and E51 in][]{pau06} are detected as 
one source in our catalog. The other 12 massive stars in~\citet{pau06} do 
not pass our detection threshold, at least partly because of the high 
background. These stars are listed in Table~\ref{t:pal_miss}. 

Among the 
13 young massive stars which do not belong to any of the three clusters
~\citep[see ][ and references therein]{mau07}, only one source is not detected
as a Paschen-$\alpha$ candidate. This source is only 1.6$\sigma_{tot}$ above 
the local strong extended Paschen-$\alpha$ emission. The line emission
in ground-based spectroscopic observations of the source may be significantly
contaminated by the nebula emission, as proposed by~\citet{cot99} .

\section{Summary}\label{s:summary}
In this paper, we have detailed the data cleaning, calibration, and
analysis procedures for our large-scale HST/NICMOS survey of the GC.
The key steps in these procedures, implemented specifically for this survey,
include: (1) the removal of the telescope and instrument effects,
particular the
 DC offsets within the four quadrants of individual
exposures and among different
position images;
 (2) the correction for the
relative and absolute astrometry of the data; and (3) the quantification
of the photometric uncertainties for sources detected with `Starfinder'.
Our main products and preliminary interpretations are as follows:
\begin{itemize}
\item We have constructed the background-subtracted, astrometry-corrected mosaics
of the net Paschen-$\alpha$ intensity as well as the F187N and F190N filter
images for the central $\sim~100\times40$ ${\rm pc}$$^2$ of our
Galaxy, providing high resolution ($\sim$0.2$\arcsec$), high fidelity
data with an average sensitivity of $\sim$ 90 $\mu${\rm Jy} ${\rm arcsecond^{-2}}$ for F187N 
and F190N and $\sim$ 130 $\mu${\rm Jy} ${\rm arcsecond^{-2}}$ for Paschen-$\alpha$.

\item We have built a catalog of $\sim~$0.6 {\rm million} point-like sources detected
in the both F187N and F190N filters. These sources contribute up to 85\% of the
total intensity observed in the F190 band. The 50\% detection limit
varies from 15.5th magnitude near Sgr A* to 17.5th magnitude in
regions of lowest stellar density. The sources should represent
predominantly evolved low-mass stars, with a much smaller component consisting of 
MS or evolved massive stars ($\gtrsim 7 M_\odot$).
A fraction of 54\% of the GC sources in the extinction range $A_{F190N} = 1.8-4.7$ tend to
be substantially brighter (intrinsically) than both foreground and background
stars with lower or higher extinction. This trend is most likely
caused by the presence of a prominent RC clump (at about 15.8th
mag). A steep extinction curve toward the \gc\ \citep{nis09} is needed
to simultaneously explain the magnitudes and colors of these RCs.

\item We have obtainted a median F187N/F190N flux ratio map, adaptively and
statistically constructed from detected source fluxes to trace the
foreground extinction of the GC at a spatial resoluition of
10$\arcsec$. This map allows for a more reliable estimation of the
F187N band stellar continuum and hence the net Paschen$-\alpha$
emission from the GC. This also provides one of the highest resolution extinction maps of the survey region to date,  although the closeness of the two filters may result in large  
systematic uncertainties in the extinction toward individual lines
of sight ($\sim$0.2 mag in K band).

\item We have presented a primary catalog of 152 Paschen-$\alpha$ emitting
candidates, plus a secondary list of 189 more tentative identifications.
These sources mostly represent evolved very massive stars with
strong optically-thin stellar winds, as partly confirmed by existing and
follow-up spectroscopic observations. In particular, the candidates 
detected first in our uniform survey are mostly located
outside the three known clusters and represent the large-scale low-intensity
star formation processes in the exreme environment of the \gc . These detections represent a significant increase in the number, and an important diversification in the location, of known young massive stars within the Galactic center. 
\end{itemize}

The data products of the survey, the catalogs and images described in
the present paper, will also be released to the public via the Legacy
Archive of the Space Telescope Science Institute. 

\section*{Acknowledgments}
We gratefully acknowledge the support of the staff at STScI
for helping in the data reduction
and analysis. We thank the referee, 
 Paco Najarro, for useful suggestions about the red clump stars and
 the extinction curve toward the GC. Support for program HST-GO-11120 was provided by NASA through a grant from the Space Telescope Science Institute, which is operated by the Association of Universities for Research in Astronomy, Inc., under NASA contract NAS 5-26555.

\appendix
\section{Global Parameter Optimization Based on Fitting to Multiple overlapping Regions}\label{s:appendix} In the main text, we have mentioned
the use of global fitting to multiple overlapping regions to optimize
parameter determinations. Such parameters could be spatial offsets
(\S\ref{s:astrometry}) or relative
background corrections (\S\ref{s:pedestal})
between adjacent images, constructed for the quadrants,
positions or orbits (we call them simply as `parts' below for simplicity).
We label the parameters as $\delta\overrightarrow{X}_{i}$, "i"
is the ID for different parts). In each case (the spatial offset or the
background correction), we minimize a specific defined global $\chi^2$
to the optimal parameters for all parts.

We first define the global $\chi^2$. For two arbitrary adjacent parts,
their best parameter differences and errors ($\overrightarrow{X}_{ij}$
and $\sigma_{ij}$, $\overrightarrow{X}_{ij}$=-$\overrightarrow{X}_{ji}$,
$\sigma_{ij}$=$\sigma_{ji}$) could be calculated from the
cross-correlation method (for the spatial offset) or the median difference
(for the background correction) through the pixel values in the
overlapping region. If two parts are not adjacent, we
set $\overrightarrow{X}_{ij}$=0 and $\sigma_{ij}$=$\infty$.
Therefore, the formula for $\chi^2$ can be expressed as
\begin{equation}
\chi^2=\sum_{i,j}\frac{(\delta\overrightarrow{X}_{i}-\delta\overrightarrow{X}_{j}+\overrightarrow{X}_{ij})^2}{\sigma_{ij}^2}
\end{equation}

Then, in order to get a global minimum for $\chi^2$, we can create
an equation array.
\begin{equation}
\frac{\partial\chi^2}{\partial\delta\overrightarrow{X}_{i}}=\sum_{j}\frac{4\times(\delta\overrightarrow{X}_{i}-\delta\overrightarrow{X}_{j}+\overrightarrow{X}_{ij})}{\sigma_{ij}^2}=0\\
\end{equation}

i.e.
$$
\left[
\begin{array}{ccccc}
  \displaystyle\sum_{j}\frac{1}{\sigma_{1j}^2} & -\displaystyle\frac{1}{\sigma_{12}^2} & ... & ... & -\displaystyle\frac{1}{\sigma_{1N}^2} \\
  -\displaystyle\frac{1}{\sigma_{21}^2} & \displaystyle\sum_{j}\frac{1}{\sigma_{2j}^2} & ... & ... & -\displaystyle\frac{1}{\sigma_{2N}^2} \\
  ... & ... & ... & ... & ... \\
  ... & ... & ... & ... & ...\\
 -\displaystyle\frac{1}{\sigma_{N1}^2} & -\displaystyle\frac{1}{\sigma_{N2}^2} & ... & ... &
 \displaystyle\sum_{j}\frac{1}{\sigma_{Nj}^2} \\
\end{array}\right] \times
\left[
\begin{array}{c}
\delta\overrightarrow{X}_{1} \\
\delta\overrightarrow{X}_{2} \\
...\\
...\\
\delta\overrightarrow{X}_{N} \\
\end{array}\right] =
\left[
\begin{array}{c}
-\displaystyle\sum_{j}\frac{\overrightarrow{X}_{1j}^2}{\sigma_{1j}^2}\\
-\displaystyle\sum_{j}\frac{\overrightarrow{X}_{2j}^2}{\sigma_{2j}^2}\\
...\\
...\\
-\displaystyle\sum_{j}\frac{\overrightarrow{X}_{Nj}^2}{\sigma_{Nj}^2}\\
\end{array}
\right]~~~~~~~~~~~~~~~~~~~~(4)
$$

$N$ is the total number of parameters (288 in the astrometry correction, see 
\S\ref{s:astrometry}, 16 and 576 in the background correction 
among quadrants and positions, see \S\ref{s:pedestal}). 
Since either in correcting the
background difference in different quadrants, positions (see
\S\ref{s:pedestal}) or relative astrometry (see
\S\ref{s:astrometry}), we always calculate the relative
difference among the parts. Therefore, the equations for the
parameters of one part (one equation in background difference
calculation and two equation in astrometry correction for
$\Delta\alpha$ 
and $\Delta \delta$ of one orbit) in the equation array should be
extra. In order to get an uniform solution, we set $\Delta
\alpha_{1}$=0 and $\Delta \delta_{1}$=0 when calculating the relative
astrometry (see \S\ref{s:astrometry}) and in the background
difference calculation between quadrants and positions (see
\S\ref{s:pedestal}), we add one more constrain that the
sum of the total background correction for all the parts should be
0. Then through solving the 2$\times (N-1)$ (astrometry correction,
the spatial shift of orbit 1 was fixed) or N (the background
difference) equations, we can obtain the parameters for the N
parts.

\begin{deluxetable}{lr}
  \tabletypesize{\footnotesize}
  \tablecaption{Survey Parameters}
  \tablewidth{0pt}
  \tablehead{
  \colhead{Parameters} &
  \colhead{Value}
  }
  \startdata
  Instrument & \nicmos\ NIC3 \\
  Total \# of Orbits & 144 \\
  Sky Coverage (arcmin$^2$) & 416  \\
  \# of Positions per Orbit & 4\\
  \# of Dither Exposures per Position & 4\\
  Field of View of Each Position & $52^\prime\times52^\prime$\\
  Filters & F187N/F190N  \\
          & (1.87$\mu$m, on-line)/(1.90$\mu$m,
  off-line) \\
  Effective wavelength & 1.8748$\mu$m/1.9003$\mu$m\\
  PHOTFNU & 43.2/40.4 $\mu$Jy sec $DN^{-1}$\\
  $F_{nu}(Vega)$ & 803.8/835.6 Jy\\
  Exposure per Filter/Position (s) & 192\\
  Readout Mode & MULTIACCUM \\
  \enddata
 \label{t:observation}
 \end{deluxetable}

\begin{deluxetable}{ccccccccccccccc}
  \tabletypesize{\scriptsize}
  \rotate
  \tablecolumns{15}
  \tablecaption{\hst /\nicmos\ \gc\ Survey Gatalog}
  \tablewidth{0pt}
  \tablehead{
  \colhead{}&
  \colhead{R.A.}&
  \colhead{Decl.}&
  \multicolumn{2}{c}{Uncertainty}&
  \colhead{}&
  \colhead{}&
  \colhead{$f_{1.87~\mu m}$}&
  \colhead{$f_{1.90~\mu m}$}&
  \colhead{$\sigma_{f_{1.87~\mu m}}$}&
  \colhead{$\sigma_{f_{1.90~\mu m}}$}&
  \colhead{}&
  \colhead{}&
  \colhead{}&
  \colhead{Data}\\
  \colhead{Source ID} &
  \colhead{(J2000.0)} &
  \colhead{(J2000.0)} &
  \colhead{(R.A.)} &
  \colhead{(Decl.)} &
  \colhead{l}&
  \colhead{b}&
  \colhead{($Jy$)}&
  \colhead{($Jy$)}&
  \colhead{($Jy$)}&
  \colhead{($Jy$)}&
  \colhead{$m_{F187N}$}&
  \colhead{$m_{F190N}$}&
  \colhead{$N_{exp}$}&
  \colhead{Quality}\\
  \colhead{(1)} &
  \colhead{(2)} &
  \colhead{(3)} &
  \colhead{(4)}&
  \colhead{(5)}&
  \colhead{(6)}&
  \colhead{(7)}&
  \colhead{(8)}&
  \colhead{(9)}&
  \colhead{(10)}&
  \colhead{(11)}&
  \colhead{(12)}&
  \colhead{(13)}\\
  }
  \startdata
 1&266.27877&-29.14364&0.02&0.03&359.76538&-0.01395&5.36&5.35&0.09&0.08&5.4&5.5&4&0\\
 2&266.42287&-28.86000&0.05&0.04&  0.07313& 0.02640&4.09&4.16&0.07&0.07&5.7&5.8&4&0\\
 3&266.28008&-29.13699&0.02&0.03&359.77165&-0.01146&3.41&3.41&0.06&0.05&5.9&6.0&4&0\\
 4&266.46723&-28.78837&0.02&0.01&  0.15452& 0.03053&2.90&3.38&0.07&0.08&6.1&6.0&2&0\\
 5&266.51105&-28.99267&0.05&0.01&  0.00003&-0.10856&3.27&3.23&0.06&0.05&6.0&6.0&4&0\\
 6&266.26167&-29.11459&0.04&0.02&359.78236& 0.01396&2.74&2.87&0.05&0.05&6.2&6.2&4&0\\
 7&266.59299&-28.74451&0.02&0.01&  0.24932&-0.04081&2.82&2.85&0.05&0.04&6.1&6.2&4&0\\
 8&266.44895&-29.05837&0.04&0.02&359.91567&-0.09639&2.73&2.78&0.05&0.04&6.2&6.2&4&0\\
 9&266.24587&-29.27417&0.02&0.04&359.63905&-0.05760&2.22&2.31&0.06&0.05&6.4&6.4&2&0\\
10&266.47656&-28.94451&0.03&0.03&  0.02546&-0.05773&2.38&2.29&0.04&0.04&6.3&6.4&4&0\\
11&266.49101&-28.98780&0.02&0.02&359.99507&-0.09106&1.86&2.11&0.03&0.03&6.6&6.5&4&0\\
12&266.38384&-28.78755&0.01&0.04&  0.11715& 0.09334&1.83&1.86&0.03&0.03&6.6&6.6&4&0\\
13&266.40656&-28.77856&0.02&0.02&  0.13519& 0.08103&1.59&1.64&0.03&0.03&6.8&6.8&4&0\\
14&266.48984&-28.74051&0.01&0.04&  0.20570& 0.03851&1.42&1.41&0.03&0.02&6.9&6.9&4&0\\
15&266.38173&-29.01626&0.01&0.02&359.92100&-0.02428&1.38&1.40&0.03&0.03&6.9&6.9&2&0\\
16&266.26673&-29.18231&0.02&0.03&359.72691&-0.02517&1.26&1.40&0.02&0.02&7.0&6.9&4&0\\
17&266.51293&-28.81531&0.03&0.03&  0.15235&-0.01769&1.27&1.30&0.02&0.02&7.0&7.0&4&0\\
18&266.24782&-29.09057&0.03&0.02&359.79653& 0.03682&1.41&1.10&0.02&0.02&6.9&7.2&4&0\\
19&266.24378&-29.25220&0.01&0.02&359.65684&-0.04457&1.05&1.06&0.02&0.02&7.2&7.2&3&0\\
20&266.40381&-29.06478&0.03&0.03&359.88964&-0.06605&1.02&1.05&0.02&0.02&7.2&7.3&4&0\\
...&...&...&...&...&...&...&...&...&...&...&...&...&...&...\\
 \enddata
\tablecomments{The full source list is published online in the future. A portion is shown here. Units of R.A. and Decl. are decimal degrees, while units of uncertainty is arcsecond. Columns 6 and 7 are in Galactic longitude and latitude in decimal degrees. The sources with the `Data Quality'=1 have nearby bad pixels, while the sources with the `Data quality'=2 are significantly contaminated by wings of bright sources (see \S~\ref{s:detect})}
 \label{t:source}
 \end{deluxetable}

\begin{deluxetable}{cccccccccccc}
  \tabletypesize{\scriptsize}
  \tablecolumns{12}
  \tablecaption{Primary Paschen-$\alpha$ emitting candidates}
  \tablewidth{0pt}
  \tablehead{
  \colhead{Source} &
  \colhead{R.A.} &
  \colhead{Decl.} &
  \colhead{} &
  \colhead{} &
  \colhead{} & 
  \colhead{} &
  \colhead{} & 
  \colhead{} &
  \colhead{} &
  \colhead{} &
  \colhead{}\\
  \colhead{ID} &
  \colhead{(J2000.0)} &
  \colhead{(J2000.0)} &
  \colhead{H}&
  \colhead{K}&
  \colhead{$m_{F190N}$} &
  \colhead{r} &
  \colhead{r-$\bar{r}$} &
  \colhead{$N_{s}$} & 
  \colhead{Counterpart} &
  \colhead{Type} &
  \colhead{Location} \\
  \colhead{(1)} &
  \colhead{(2)} &
  \colhead{(3)} &
  \colhead{(4)}&
  \colhead{(5)}&
  \colhead{(6)}&
  \colhead{(7)}&
  \colhead{(8)}&
  \colhead{(9)}&
  \colhead{(10)}&
  \colhead{(11)}&
  \colhead{(12)}\\
  }
  \startdata
  1&266.62478&-28.78001&    12.8&    12.6&12.6& 1.1&0.15& 5.6&                    &                    &F\\
  2&266.59921&-28.80299&    13.3&    11.4&12.1& 2.6&1.64&26.7&          Mau10c\_17&                WN5b&F\\
  3&266.51501&-28.78606&    14.1&    12.3&13.0& 1.1&0.13& 4.9&                    &                    &F\\
  4&266.55606&-28.81641&    11.5&     9.8&10.5& 1.2&0.31&10.3&             FQ\_381&                 OBI&Q\\
  5&266.55408&-28.82014&    14.1&    12.2&13.8& 1.1&0.17& 5.9&                    &                    &Q\\
  6&266.56301&-28.82693&    10.5&    10.4& 9.6& 1.5&0.53&15.1&     Lie\_71,FQ\_241&                 WN9&Q\\
  7&266.56643&-28.82714&10.6$^a$& 8.9$^a$&10.2& 1.5&0.55&15.4&     Lie\_67,FQ\_240&                 WN9&Q\\
  8&266.56293&-28.82480&    11.2&     9.8&10.2& 1.2&0.23& 8.1&   Lie\_110,FQ\_270S&   O6-8 I f (Of/WN?)&Q\\
  9&266.56312&-28.82568&    11.1&    10.6&10.3& 1.1&0.17& 6.2&  Lie\_96,Mau10a\_19&          O6-8 I f e&Q\\
 10&266.56304&-28.82631&    11.6&     9.9&10.6& 1.1&0.18& 6.7&     Lie\_77,FQ\_278&         O6-8 I f eq&Q\\
 11&266.56896&-28.82547&    11.7&    10.2&10.9& 1.5&0.52&14.9&     Lie\_99,FQ\_256&                 WN9&Q\\
 12&266.56323&-28.82821&    13.2&    11.3&12.2& 2.7&1.74&26.8&             Lie\_34&                 WC8&Q\\
 13&266.56316&-28.82761&        &        &12.2& 2.3&1.33&23.8&             Lie\_47&                 WC8&Q\\
 14&266.58167&-28.83606&        &        &15.2& 1.2&0.30& 4.8&                    &                    &F\\
 15&266.51338&-28.81622&    13.2&    11.7&12.4& 1.1&0.17& 6.4&                    &                    &F\\
 16&266.47853&-28.78699&    14.6&    13.1&13.6& 1.1&0.20& 6.8&                    &                    &F\\
 17&266.46028&-28.82543&    12.5&    10.6&11.3& 2.0&1.05&22.1&     FA\_5,Blu01\_22&              WN8-9h&F\\
 18&266.45712&-28.82372&    13.0&    10.8&11.7& 1.7&0.79&19.3&FA\_2,Blu01\_34,Mau1&              WN8-9h&A\\
 19&266.45255&-28.82840&    13.6&    11.1&12.1& 1.9&0.97&21.4&          Mau10c\_11&              WN8-9h&F\\
 20&266.45865&-28.82393&    12.8&    11.0&11.7& 1.2&0.29& 9.7&    FA\_10,Blu01\_30&              O4-6If&A\\
 21&266.45895&-28.82411&    13.5&    11.6&12.5& 1.2&0.26& 8.5&    FA\_17,Blu01\_29&                    &A\\
 22&266.47249&-28.82693&    12.5&    11.0&11.6& 1.3&0.31&10.5&          Mau10c\_12&              WN8-9h&F\\
 23&266.54169&-28.92566&    12.3&    10.8&11.4& 1.3&0.33&10.8&          Mau10c\_15&              WN8-9h&F\\
 24&266.50251&-28.90761&    15.2&    13.6&14.9& 1.2&0.25& 5.7&                    &                    &F\\
 25&266.49295&-28.87223&        &        &14.4& 1.2&0.30& 8.5&                    &                    &F\\
 26&266.49541&-28.89392&    15.5&    13.6&15.7& 1.4&0.42& 4.6&                    &                    &F\\
 27&266.48141&-28.90196&    16.2&    15.3&15.3& 1.4&0.45& 7.5&                    &                    &F\\
 28&266.49067&-28.91267&    13.4&    11.4&12.2& 1.9&1.00&21.5&                 Ho2&               WC8-9&F\\
 29&266.53998&-28.95375&    16.5&    14.4&14.9& 1.1&0.17& 4.9&                    &                    &F\\
 30&266.53377&-28.97338&    12.6&    10.5&11.0& 1.1&0.18& 6.6&                    &                    &F\\
 31&266.52612&-28.98747&    13.6&    11.7&12.3& 1.1&0.12& 4.8&                    &                    &F\\
 32&266.37803&-28.87674&        &        &17.1& 1.6&0.66& 5.0&                    &                    &F\\
 33&266.33953&-28.86082&        &        &14.2& 1.2&0.26& 8.1&                    &                    &F\\
 34&266.46061&-28.95726&    12.9&    11.3&12.0& 2.6&1.67&26.9&          Mau10c\_19&                 WC9&F\\
 35&266.36926&-28.93473&    11.5&     9.7&10.6& 1.4&0.46&10.8&    Cot\_4,Mau10a\_7&                  Of&F\\
 36&266.38126&-28.95466&    12.7&    11.4&11.8& 1.1&0.17& 6.4&           Mau10c\_6&               O4-6I&F\\
 37&266.44762&-29.04884&        &        &15.1& 1.2&0.22& 4.6&                    &                    &F\\
 38&266.40831&-29.02624& 9.6$^a$& 8.9$^a$& 9.1& 1.1&0.13& 5.2&           Mau10c\_7&               O4-6I&F\\
 39&266.34458&-28.97893&    15.2&    12.2&13.4& 2.3&1.41&24.8&           Mau10a\_6&              WN5-6b&F\\
 40&266.35029&-29.01633&10.3$^a$& 8.8$^a$& 9.7& 1.2&0.21& 7.5&           Mau10c\_5&             B0I-B2I&F\\
 41&266.40788&-29.10817&    13.1&    11.5&12.1& 1.1&0.15& 5.9&                    &                    &F\\
 42&266.38541&-29.08277&    14.6&    12.0&13.0& 1.8&0.88&20.1&           Mau10c\_8&                 WC9&F\\
 43&266.30814&-29.07728&    15.2&    13.6&15.6& 1.3&0.37& 5.7&                    &                    &F\\
 44&266.25555&-29.04208&    14.8&    11.9&13.1& 1.1&0.18& 6.6&                    &                    &F\\
 45&266.25250&-29.10650&    16.6&    14.8&15.8& 1.2&0.26& 5.2&                    &                    &F\\
 46&266.28706&-29.11563&    14.9&    13.3&13.9& 1.3&0.38&11.1&                    &                    &F\\
 47&266.31271&-29.15230&        &        &14.9& 1.1&0.17& 4.7&                    &                    &F\\
 48&266.34449&-29.18235&        &        &15.7& 1.3&0.34& 6.0&                    &                    &F\\
 49&266.34120&-29.19988&    14.7&    12.7&13.4& 1.9&0.94&20.5&           Mau10c\_3&                 WC9&F\\
 50&266.26206&-29.14994&        &        &10.6& 1.1&0.15& 5.8&           Mau10a\_1&             O9I-B0I&F\\
 51&266.26160&-29.13144&    16.1&    14.5&14.6& 1.1&0.18& 5.7&                    &                    &F\\
 52&266.22865&-29.11919&    16.6&    14.8&15.1& 1.1&0.17& 5.0&                    &                    &F\\
 53&266.27944&-29.20018&    13.5&    11.1&12.1& 1.3&0.34&11.1&           Mau10c\_2&               WC9?d&F\\
 54&266.24580&-29.22798&    15.9&    13.8&15.0& 1.9&0.95&17.2&                    &                    &F\\
 55&266.24935&-29.26869&12.6$^a$&11.9$^a$&14.6& 1.1&0.17& 5.1&                    &                    &F\\
 56&266.61499&-28.76995&    11.3&     9.5&10.2& 1.3&0.35&11.3&          Mau10c\_18&                  OI&F\\
 57&266.62457&-28.77776&    14.7&    12.6&13.6& 1.3&0.38&11.6&                    &                    &F\\
 58&266.63270&-28.77975&    12.6&    11.6&11.9& 1.2&0.34& 5.8&                    &                    &F\\
 59&266.57304&-28.82467&11.8$^a$&10.2$^a$&11.0& 1.4&0.46& 9.6&             FQ\_274&                 WN9&Q\\
 60&266.57294&-28.82181&    13.4&    11.4&12.2& 2.2&1.27&24.1&             FQ\_309&                 WC8&Q\\
 61&266.48245&-28.74278&    15.0&    13.3&14.0& 1.3&0.35&10.5&                    &                    &F\\
 62&266.55855&-28.82124&    12.3&    10.5&11.3& 2.2&1.24&24.0&    Lie\_158,FQ\_320&                 WN9&Q\\
 63&266.55437&-28.82364&    12.9&    10.5&11.5& 1.4&0.49&14.4&                 Ho3&               WC8-9&Q\\
 64&266.54641&-28.81830&    13.4&    11.6&12.3& 2.6&1.61&26.5&            FQ\_353E&                 WN6&F\\
 65&266.55773&-28.81403&    12.7&    10.9&11.7& 1.1&0.21& 7.6&             FQ\_406&                    &Q\\
 66&266.56483&-28.83833&    13.2&    11.2&12.2& 2.0&1.01&15.3&              FQ\_76&                 WC9&Q\\
 67&266.56301&-28.83910&        &        &15.6& 1.4&0.41& 4.6&                    &                    &Q\\
 68&266.56352&-28.83428& 8.9$^a$& 7.3$^a$& 7.3& 1.2&0.25& 7.6&             FQ\_134&                 LBV&Q\\
 69&266.56452&-28.82226&    10.8&     9.2& 9.9& 1.1&0.12& 4.5&    Lie\_146,FQ\_307&           O6-8 I f?&Q\\
 70&266.56973&-28.83090&    11.9&    10.2&11.0& 1.1&0.18& 6.4&              Lie\_1&           O3-8 I fe&Q\\
 71&266.55895&-28.82645&    12.9&    10.3&11.5& 1.4&0.45&13.3&             Lie\_76&                WC9d&Q\\
 72&266.56676&-28.82268&    12.1&    10.5&11.3& 1.1&0.14& 5.1&    Lie\_143,FQ\_301&             O7-B0 I&Q\\
 73&266.56173&-28.83344&    13.2&    10.5&11.9& 1.2&0.22& 5.3&             FQ\_151&                 WC8&Q\\
 74&266.55722&-28.82803&        &        &15.2& 1.4&0.46& 7.7&                    &                    &Q\\
 75&266.57118&-28.85862&    12.1&    10.5&11.3& 1.3&0.36&11.2&   M07\_2,Mau10a\_22&               O6If+&F\\
 76&266.58645&-28.87528&    13.7&    12.6&13.1& 1.1&0.14& 4.6&                    &                    &F\\
 77&266.57310&-28.88431&    12.1&    10.5&11.1& 1.5&0.51&14.6&          Mau10c\_16&              WN8-9h&F\\
 78&266.45091&-28.79055&    14.6&    12.7&13.6& 1.1&0.15& 5.6&                    &                    &F\\
 79&266.46449&-28.82372&    13.2&    11.2&11.8& 1.5&0.61&11.8&            Blu01\_1&                 WN7&F\\
 80&266.46014&-28.82276&    11.6&     9.9&10.6& 2.0&1.09&22.3&FA\_6,Blu01\_23,Mau1&              WN8-9h&A\\
 81&266.46075&-28.82145&    11.7&    10.1&10.8& 2.1&1.21&20.3&     FA\_4,Blu01\_17&              WN7-8h&A\\
 82&266.46182&-28.82389&    12.0&    10.1&10.9& 2.1&1.14&22.9&      FA\_3,Blu01\_3&              WN8-9h&A\\
 83&266.46035&-28.82199&    13.6&    12.0&10.7& 1.8&0.82&18.8&FA\_7,Blu01\_21,Mau1&              WN8-9h&A\\
 84&266.46001&-28.82246&    12.0&    10.1&11.1& 2.0&1.04&21.6&     FA\_8,Blu01\_24&              WN8-9h&A\\
 85&266.45925&-28.82274&    12.0&    10.1&11.0& 1.7&0.79&18.9&     FA\_1,Blu01\_28&              WN8-9h&A\\
 86&266.45948&-28.81983&    12.2&    10.5&11.0& 1.5&0.56&11.2&FA\_9,Blu01\_26,Mau1&              WN8-9h&A\\
 87&266.45954&-28.82136&10.0$^a$& 9.6$^a$&11.4& 1.8&0.86&19.5&    FA\_12,Blu01\_25&              WN7-8h&A\\
 88&266.46121&-28.82284&    12.4&    10.8&11.5& 1.6&0.69&17.3&    FA\_14,Blu01\_12&              WN8-9h&A\\
 89&266.46151&-28.82118&    12.2&    10.8&11.4& 1.2&0.28& 6.6&     FA\_15,Blu01\_8&              O4-6If&A\\
 90&266.46057&-28.82231&    12.5&    10.9&11.7& 1.4&0.50&13.3&    FA\_16,Blu01\_19&              WN8-9h&A\\
 91&266.48072&-28.85726&    12.5&    11.0&11.5& 2.5&1.61&18.8&   M07\_1,Mau10a\_16&              WN5-6b&F\\
 92&266.52345&-28.85881& 9.2$^a$& 7.5$^a$& 7.4& 1.3&0.38&11.9&              Mau10b&                 LBV&F\\
 93&266.51671&-28.89814&        &        &15.6& 1.2&0.30& 5.5&                    &                    &F\\
 94&266.51080&-28.90388&    13.5&    11.6&12.5& 2.4&1.42&21.8&          Mau10c\_14&                 WC9&F\\
 95&266.49765&-28.88075&    12.1&    10.5&11.2& 1.1&0.12& 4.8&                    &                    &F\\
 96&266.45238&-28.83491&    13.3&    11.0&11.8& 2.0&1.10&22.8&              Cot\_1&            Ofpe/WN9&F\\
 97&266.44891&-28.84692&12.2$^a$&10.7$^a$&11.5& 1.2&0.25& 7.6&                    &                    &F\\
 98&266.42197&-28.86325&    11.6&     9.9&10.6& 1.2&0.27& 6.6&           Mau10c\_9&             O4-6If+&F\\
 99&266.41118&-28.86958&    16.4&    13.9&15.1& 1.9&0.95&17.4&                    &                    &F\\
100&266.42634&-28.87976&    11.7&    10.1&10.9& 1.3&0.37& 8.2&          Mau10c\_10&             O4-6If+&F\\
101&266.42704&-28.88140&    13.2&    11.2&12.1& 1.9&0.98&21.0&                 Ho1&               WC8-9&F\\
102&266.43349&-28.88800&    13.3&    12.4&12.8& 1.2&0.25& 8.8&                    &                    &F\\
103&266.50689&-28.92091&    10.7&     9.1& 9.7& 1.2&0.28& 9.7&          Mau10c\_13&                  OI&F\\
104&266.53042&-28.95485&        &        &16.3& 1.5&0.52& 4.5&                    &                    &F\\
105&266.47070&-28.92692&    14.9&    13.4&14.0& 1.1&0.19& 6.2&                    &                    &F\\
106&266.46776&-28.94615&    16.1&    14.5&14.7& 1.3&0.32& 8.1&                    &                    &F\\
107&266.41391&-28.88919&    11.8&    10.2&10.9& 1.4&0.43&11.4&              Cot\_5&                B[e]&F\\
108&266.44602&-28.94612&    15.5&    14.0&14.7& 1.2&0.27& 6.4&                    &                    &F\\
109&266.46088&-28.98877&    12.5&    10.9&11.6& 1.9&1.01&19.2&   Cot\_2,Mau10a\_15&                 WN7&F\\
110&266.45393&-28.98382&    15.7&    14.3&15.0& 1.1&0.20& 4.7&                    &                    &F\\
111&266.40061&-28.94403&    12.3&    10.4&11.1& 1.8&0.85&20.0& Mik06\_01,Mau10a\_9&              WN8-9h&F\\
112&266.40753&-28.95450&    13.1&    10.7&12.0& 1.4&0.49&13.8&              Cot\_6&                B[e]&F\\
113&266.39038&-28.96381&    14.2&    12.2&12.9& 1.1&0.19& 6.9&                    &                    &F\\
114&266.38658&-28.93794&    12.1&    10.7&11.3& 1.2&0.21& 7.1&           Mau10a\_8&               O4-6I&F\\
115&266.32590&-28.89079&    13.8&    11.9&12.9& 1.1&0.16& 5.1&                    &                    &F\\
116&266.29639&-28.91955&    13.0&    12.8&12.8& 1.1&0.16& 6.0&                    &                    &F\\
117&266.41477&-29.00973&    12.7&    10.3&11.6& 2.9&2.00&20.0&                 E79&            Ofpe/WN9&C\\
118&266.41443&-29.00881&        &        &12.6& 1.7&0.79&12.8&                 E74&                 WN8&C\\
119&266.41377&-29.00853&        &        &13.5& 2.0&1.09&10.6&                 E81&                 WN7&C\\
120&266.41410&-29.00929&        &        &13.7& 1.8&0.90&10.7&                 E82&               WC8/9&C\\
121&266.41777&-29.00756&        &        &10.0& 1.2&0.27& 5.2&                 E39&            Ofpe/WN9&C\\
122&266.41586&-29.00830&    11.6&     9.1&10.9& 1.6&0.68& 8.6&             E51+E48&             WN8+WC9&C\\
123&266.41728&-29.00458&        &        &12.4& 2.0&1.08&16.5&                 E88&               WN8/9&C\\
124&266.41635&-29.00504&    13.3&    11.5&12.6& 1.7&0.74&12.7&                 E83&             WN8/WC9&C\\
125&266.41555&-29.00739&        &        &12.7& 1.3&0.38& 5.8&                 E56&            Ofpe/WN9&C\\
126&266.41560&-29.00646&    14.2&    12.0&12.9& 1.5&0.60&11.9&                 E66&                 WN8&C\\
127&266.41714&-29.00621&        &        &13.0& 1.2&0.31& 4.7&                 E71&             WC8/9 ?&C\\
128&266.41608&-29.00616&        &        &14.3& 1.7&0.76& 6.3&                 E68&                 WC9&C\\
129&266.41717&-29.00767&        &        &10.6& 1.2&0.28& 5.2&                 E20&            Ofpe/WN9&C\\
130&266.41777&-29.00814&        &        &13.0& 2.4&1.49& 6.2&                 E40&               WN5/6&C\\
131&266.41984&-29.00772&        &        &13.9& 1.8&0.88& 7.8&                 E78&                 WC9&C\\
132&266.41774&-29.00936&    13.6&    11.1&13.0& 1.9&0.98& 5.4&                 E65&                 WN8&C\\
133&266.40359&-29.02154&        &        &16.0& 2.1&1.15& 5.5&                    &                    &F\\
134&266.31975&-28.97363&    13.1&    11.1&12.1& 1.8&0.84&13.8&           Mau10a\_4&              WN7-8h&F\\
135&266.32402&-28.97220&    16.1&    14.1&15.1& 1.3&0.33& 7.3&                    &                    &F\\
136&266.36759&-29.05754&    14.8&    12.7&13.3& 1.1&0.16& 5.8&                    &                    &F\\
137&266.31749&-29.05437& 9.2$^a$& 7.9$^a$& 8.3& 1.6&0.68& 8.7&Muno06\_01,Mau10a\_3&            Ofpe/WN9&F\\
138&266.33870&-29.13177&        &        &14.5& 1.1&0.20& 6.4&                    &                    &F\\
139&266.31898&-29.09621&    15.4&    13.5&14.2& 1.1&0.20& 6.6&                    &                    &F\\
140&266.24782&-29.09057& 7.6$^a$& 7.0$^a$& 7.2& 1.3&0.35&11.3&           Mau10c\_1&             B0I-B2I&F\\
141&266.28870&-29.13783&    15.3&    11.7&13.2& 1.1&0.18& 6.7&                    &                    &F\\
142&266.33105&-29.17609&    15.9&    14.8&15.0& 1.2&0.22& 5.6&                    &                    &F\\
143&266.34964&-29.17353&        &        &16.2& 1.3&0.36& 5.0&                    &                    &F\\
144&266.30924&-29.19493&        &        &15.6& 1.3&0.38& 7.1&                    &                    &F\\
145&266.32051&-29.20505& 8.6$^a$& 8.0$^a$& 8.2& 1.1&0.12& 4.7&                    &                    &F\\
146&266.22471&-29.09574&        &        &14.6& 2.0&1.10&11.2&                    &                    &F\\
147&266.28741&-29.20498&    12.6&    11.1&11.6& 2.2&1.20&23.4&           Mau10a\_2&                 WN7&F\\
148&266.30920&-29.26002&    16.2&    14.7&15.5& 1.2&0.25& 4.7&                    &                    &F\\
149&266.31115&-29.25314&    13.1&    12.4&12.8& 1.5&0.53&15.0&                    &                    &F\\
150&266.31271&-29.24339&    14.5&    12.7&13.7& 1.6&0.69&16.5&                    &                    &F\\
151&266.29083&-29.23696&    13.0&    11.1&12.1& 1.5&0.53&10.6&           Mau10c\_4&               WC9?d&F\\
152&266.24459&-29.25164&        &        &15.1& 1.2&0.29& 4.7&                    &                    &F\\
\enddata
 \label{t:pal_sou}
\tablecomments{Units of R.A. and Decl. are decimal degrees. H and K band magnitudes are mainly from the SIRIUS catalog~\citep{nis06}, while the superscript `a' indicates the magnitudes are from the 2MASS catalog (see \S~\ref{s:dis}). The ground-based spetroscopically identified counterparts of our Paschen-$\alpha$ emitting sources and their types are from:~\citet{fig99a,fig02,lie09,mar08,blu01,cot99,hom03,mun06,mik06,mau07,mau10a,mau10b,mau10c}. The `Location' column divides the sources into four groups: the sources inside the three clusters (`Q': Quintuplet, `A': Arches, `C': Center) and filed sources outside the clusters (`F', see \S~\ref{s:dis})}
 \end{deluxetable}

\begin{deluxetable}{ccccccccccccc}
  \tabletypesize{\scriptsize}
  \tablecolumns{13}
  \tablecaption{Secondary Paschen-$\alpha$ emitting candidates}
  \tablewidth{0pt}
  \tablehead{
  \colhead{Source} &
  \colhead{R.A.} &
  \colhead{Decl.} &
  \colhead{} &
  \colhead{} &
  \colhead{} & 
  \colhead{} &
  \colhead{} & 
  \colhead{} &
  \colhead{} &
  \colhead{} &
  \colhead{} &
  \colhead{Problem}\\
  \colhead{ID} &
  \colhead{(J2000.0)} &
  \colhead{(J2000.0)} &
  \colhead{H}&
  \colhead{K}&
  \colhead{$m_{F190N}$} &
  \colhead{r} &
  \colhead{r-$\bar{r}$} &
  \colhead{$N_{s}$} & 
  \colhead{Counterpart} &
  \colhead{Type} &
  \colhead{Location} & 
  \colhead{Index}\\
  \colhead{(1)} &
  \colhead{(2)} &
  \colhead{(3)} &
  \colhead{(4)}&
  \colhead{(5)}&
  \colhead{(6)}&
  \colhead{(7)}&
  \colhead{(8)}&
  \colhead{(9)}&
  \colhead{(10)}&
  \colhead{(11)}&
  \colhead{(12)}&
  \colhead{(13)}\\
  }
  \startdata
  1&266.57112&-28.81928&        &        &17.0& 1.9&0.93& 5.1&                    &                    &Q&1\\
  2&266.54507&-28.84973&        &        &15.1& 1.6&0.68&13.1&                    &                    &F&1\\
  3&266.46006&-28.82544&        &        &15.4& 1.1&0.24& 4.7&     FA\_5,Blu01\_22&              WN8-9h&F&3\\
  4&266.45999&-28.82534&        &        &15.4& 1.2&0.26& 5.4&     FA\_5,Blu01\_22&              WN8-9h&F&3\\
  5&266.52087&-28.88455&        &        &16.1& 1.6&0.64& 6.4&                    &                    &F&1\\
  6&266.41235&-28.80568&        &        &16.0& 1.3&0.36& 4.6&                    &                    &F&1\\
  7&266.49792&-28.96117&        &        &15.4& 1.4&0.45& 8.3&                    &                    &F&1\\
  8&266.41772&-28.94261&        &        &15.4& 1.3&0.38& 7.0&                    &                    &F&3\\
  9&266.44740&-28.95131&        &        &15.6& 1.7&0.73&10.9&                    &                    &F&1\\
 10&266.46068&-28.95709&        &        &15.4& 1.3&0.33& 5.1&          Mau10c\_19&                 WC9&F&3\\
 11&266.47456&-29.00668&        &        &15.7& 1.2&0.30& 4.9&                    &                    &F&1\\
 12&266.44583&-28.98711&        &        &13.9& 1.1&0.16& 5.5&                    &                    &F&1\\
 13&266.40510&-28.97124&        &        &15.7& 1.6&0.69& 9.7&                    &                    &F&1\\
 14&266.32141&-28.93775&        &        &16.4& 1.8&0.89& 7.0&                    &                    &F&1\\
 15&266.40384&-29.00604&        &        &13.6& 2.1&1.14&16.7&                    &                    &F&1\\
 16&266.35968&-29.08454&        &        &15.4& 1.8&0.85&10.0&                    &                    &F&1\\
 17&266.32775&-29.05249&        &        &12.8& 1.2&0.27& 9.0&                    &                    &F&1\\
 18&266.25504&-29.04298&        &        &15.2& 1.4&0.49& 9.4&                    &                    &F&2\\
 19&266.29735&-29.18157&        &        &15.8& 1.4&0.40& 5.9&                    &                    &F&1\\
 20&266.25443&-29.11684&        &        &16.6& 1.6&0.69& 4.8&                    &                    &F&1\\
 21&266.26418&-29.22752&        &        &14.1& 1.5&0.51&12.4&                    &                    &F&1\\
 22&266.25012&-29.28868&        &        &16.2& 1.4&0.43& 4.9&                    &                    &F&1\\
 23&266.58636&-28.75999&        &        &15.8& 1.6&0.69& 6.9&                    &                    &F&3\\
 24&266.59539&-28.75383&        &        &16.2& 1.6&0.71& 6.7&                    &                    &F&1\\
 25&266.59311&-28.83139&        &        &15.2& 1.2&0.21& 5.1&                    &                    &F&2\\
 26&266.57431&-28.80947&        &        & 7.5& 1.2&0.23& 5.9&                    &                    &F&3\\
 27&266.55059&-28.79384&        &        &16.1& 1.4&0.45& 4.8&                    &                    &F&1\\
 28&266.52266&-28.77663&        &        &15.4& 1.5&0.51& 8.4&                    &                    &F&1\\
 29&266.46738&-28.72721&        &        &14.6& 1.2&0.25& 6.9&                    &                    &F&1\\
 30&266.51357&-28.84118&        &        &14.6& 1.5&0.56&11.2&                    &                    &F&1\\
 31&266.46091&-28.82198&        &        &12.2& 1.1&0.19& 4.7&    FA\_27,Blu01\_16&                    &A&2\\
 32&266.37506&-28.78707&        &        &13.8& 1.5&0.53&12.3&                    &                    &F&1\\
 33&266.49000&-28.90671&        &        &11.2& 1.1&0.17& 6.3&                    &                    &F&1\\
 34&266.49732&-28.91171&        &        &15.7& 1.3&0.40& 4.8&                    &                    &F&1\\
 35&266.45853&-28.92780&        &        &15.4& 1.5&0.53& 8.4&                    &                    &F&2\\
 36&266.41042&-28.88971&        &        &11.7& 1.1&0.17& 4.5&                    &                    &F&3\\
 37&266.44453&-28.99359&        &        &15.5& 1.3&0.32& 4.5&                    &                    &F&3\\
 38&266.31580&-28.93428&        &        &15.0& 1.1&0.16& 4.7&                    &                    &F&3\\
 39&266.37321&-28.99285&        &        &14.7& 1.2&0.29& 6.8&                    &                    &F&2\\
 40&266.27703&-29.01534&        &        &15.3& 1.3&0.37& 7.1&                    &                    &F&1\\
 41&266.28890&-29.03190&        &        &15.8& 1.9&0.93&10.5&                    &                    &F&1\\
 42&266.33596&-29.13619&        &        &15.7& 1.5&0.53& 6.7&                    &                    &F&1\\
 43&266.26707&-29.09698&        &        &16.0& 1.6&0.67& 6.5&                    &                    &F&1\\
 44&266.27018&-29.11314&        &        &13.9& 1.1&0.18& 4.9&                    &                    &F&3\\
 45&266.26230&-29.17227&        &        &15.9& 1.4&0.43& 6.8&                    &                    &F&1\\
 46&266.61970&-28.77655&        &        &17.0& 1.7&0.73& 3.7&                    &                    &F&1\\
 47&266.62947&-28.77061&    13.8&    12.2&12.8& 1.0&0.08& 3.5&                    &                    &F&0\\
 48&266.63499&-28.77076&    14.2&    13.3&13.7& 1.0&0.09& 3.6&                    &                    &F&0\\
 49&266.63035&-28.77189&    15.5&    14.4&14.9& 1.0&0.10& 3.6&                    &                    &F&0\\
 50&266.60087&-28.75860&    11.5&    11.1&11.3& 1.0&0.09& 3.6&                    &                    &F&0\\
 51&266.58243&-28.77181&    14.4&    14.2&14.3& 1.0&0.10& 3.8&                    &                    &F&0\\
 52&266.60124&-28.78142&    13.1&    12.1&12.4& 1.0&0.09& 3.9&                    &                    &F&0\\
 53&266.62200&-28.79915&    15.3&    15.2&15.2& 1.0&0.11& 3.5&                    &                    &F&0\\
 54&266.57011&-28.79874&    15.1&    14.9&14.8& 1.1&0.12& 3.7&                    &                    &F&0\\
 55&266.54139&-28.77251&        &        &16.0& 1.2&0.24& 3.7&                    &                    &F&0\\
 56&266.55458&-28.81911&        &        &14.4& 1.1&0.13& 3.8&                    &                    &Q&0\\
 57&266.56037&-28.82746&    13.4&    11.6&12.5& 1.0&0.09& 3.6&             Lie\_54&        O7-9 I-II f?&Q&0\\
 58&266.54031&-28.86550&        &        &15.4& 1.1&0.19& 4.0&                    &                    &F&1\\
 59&266.50682&-28.83333&        &        &14.0& 1.1&0.11& 3.8&                    &                    &F&0\\
 60&266.40847&-28.76901&    16.3&    14.6&15.3& 1.1&0.15& 3.8&                    &                    &F&0\\
 61&266.45859&-28.82313&    13.3&    11.5&12.1& 1.0&0.12& 4.5&    FA\_13,Blu01\_31&                    &A&0\\
 62&266.45762&-28.82503&    14.9&    13.3&14.0& 1.0&0.10& 3.7&             FA\_107&                    &F&0\\
 63&266.51118&-28.85685&        &        &15.9& 1.2&0.30& 3.8&                    &                    &F&0\\
 64&266.52091&-28.86852&    15.9&    14.5&14.9& 1.1&0.17& 4.5&                    &                    &F&0\\
 65&266.54126&-28.90324&    15.8&    14.1&14.6& 1.1&0.13& 3.9&                    &                    &F&0\\
 66&266.41645&-28.81262&        &        &16.0& 1.2&0.32& 3.6&                    &                    &F&1\\
 67&266.37114&-28.82373&        &        &16.8& 1.5&0.54& 4.1&                    &                    &F&0\\
 68&266.38810&-28.82180&    14.7&    14.7&14.5& 1.0&0.10& 3.6&                    &                    &F&0\\
 69&266.46735&-28.91708&    17.4&    14.6&15.0& 1.1&0.14& 3.9&                    &                    &F&0\\
 70&266.51105&-28.99267& 6.4$^a$& 6.0$^a$& 6.0& 1.0&0.09& 3.7&                    &                    &F&0\\
 71&266.49376&-28.97168&    12.7&    12.5&12.6& 1.0&0.09& 3.6&                    &                    &F&0\\
 72&266.40149&-28.88929&    14.8&    14.2&14.7& 1.0&0.11& 3.6&                    &                    &F&0\\
 73&266.37635&-28.86860&    13.4&    11.9&12.5& 1.0&0.10& 3.9&                    &                    &F&0\\
 74&266.44592&-28.95423&        &        &15.8& 1.2&0.28& 4.3&                    &                    &F&1\\
 75&266.46293&-28.98092&        &        &16.1& 1.3&0.39& 4.4&                    &                    &F&0\\
 76&266.48508&-28.97955&13.4$^a$&12.7$^a$&13.5& 1.0&0.09& 3.6&                    &                    &F&0\\
 77&266.47957&-28.98002&15.3$^a$&14.5$^a$&14.4& 1.0&0.10& 3.5&                    &                    &F&0\\
 78&266.47233&-29.01364&    16.7&    15.4&15.7& 1.1&0.18& 4.0&                    &                    &F&0\\
 79&266.48204&-29.02302&    16.2&    14.3&15.4& 1.1&0.17& 3.6&                    &                    &F&0\\
 80&266.43484&-29.00002&        &        &14.9& 1.2&0.24& 3.6&                    &                    &F&2\\
 81&266.42083&-28.96982&        &        &15.4& 1.1&0.18& 3.5&                    &                    &F&0\\
 82&266.43805&-28.97761&        &        &16.0& 1.4&0.46& 3.9&                    &                    &F&1\\
 83&266.38329&-28.95134&    14.4&    14.4&14.4& 1.0&0.11& 4.1&                    &                    &F&0\\
 84&266.38153&-28.95012&    15.3&    15.1&15.2& 1.0&0.13& 4.0&                    &                    &F&0\\
 85&266.35512&-28.90918&        &        &16.2& 1.2&0.27& 3.6&                    &                    &F&0\\
 86&266.31157&-28.90497&     9.5&     9.2& 9.4& 1.0&0.09& 3.8&                    &                    &F&0\\
 87&266.38100&-28.99110&        &        &16.3& 1.2&0.29& 3.5&                    &                    &F&0\\
 88&266.43206&-29.04636&    12.6&    11.3&11.8& 1.0&0.10& 4.0&                    &                    &F&0\\
 89&266.40635&-29.02650&    10.0&     9.7& 9.7& 1.0&0.09& 3.5&                    &                    &F&0\\
 90&266.40067&-29.02766&        &        &15.2& 1.2&0.25& 4.2&                    &                    &F&0\\
 91&266.39380&-29.04020&    15.5&    14.0&14.8& 1.1&0.19& 4.1&                    &                    &F&0\\
 92&266.41532&-29.03935& 8.6$^a$& 8.3$^a$& 8.5& 1.0&0.09& 3.7&                    &                    &F&0\\
 93&266.41875&-29.03449&    14.2&    13.9&14.4& 1.0&0.11& 3.7&                    &                    &F&0\\
 94&266.37951&-29.01024&        &        &15.4& 1.2&0.28& 4.2&                    &                    &F&1\\
 95&266.29619&-28.96369&    10.0&     9.8& 9.9& 1.0&0.09& 3.5&                    &                    &F&0\\
 96&266.38314&-29.10544&        &        &16.2& 1.2&0.22& 3.6&                    &                    &F&0\\
 97&266.39968&-29.09183& 8.8$^a$& 8.5$^a$& 8.7& 1.0&0.09& 3.6&                    &                    &F&0\\
 98&266.30601&-29.01895&        &        &16.0& 1.2&0.28& 4.4&                    &                    &F&0\\
 99&266.29804&-29.04358&    14.3&    13.8&14.3& 1.0&0.10& 3.9&                    &                    &F&0\\
100&266.29195&-29.06206&        &        &16.3& 1.4&0.43& 3.8&                    &                    &F&1\\
101&266.35120&-29.09667&    12.6&    12.2&12.3& 1.0&0.09& 3.7&                    &                    &F&0\\
102&266.38987&-29.12074&        &        &15.2& 1.1&0.19& 3.7&                    &                    &F&0\\
103&266.33147&-29.13788&    12.8&    10.8&11.5& 1.0&0.09& 3.8&                    &                    &F&0\\
104&266.33271&-29.12943&        &        &14.3& 1.1&0.13& 4.4&                    &                    &F&0\\
105&266.31626&-29.10374&        &        &15.5& 1.1&0.19& 3.9&                    &                    &F&2\\
106&266.33242&-29.11555&    15.3&    13.5&14.1& 1.1&0.11& 3.8&                    &                    &F&0\\
107&266.33594&-29.11435&        &        &16.0& 1.3&0.33& 3.9&                    &                    &F&1\\
108&266.27558&-29.10560&        &        &15.9& 1.2&0.24& 4.3&                    &                    &F&0\\
109&266.32970&-29.17216&    15.4&    14.7&14.7& 1.1&0.11& 3.5&                    &                    &F&0\\
110&266.26474&-29.13603&    15.3&    13.8&14.4& 1.1&0.11& 3.7&                    &                    &F&0\\
111&266.29026&-29.19196&        &        &15.6& 1.1&0.21& 4.0&                    &                    &F&0\\
112&266.23740&-29.19839&        &        &15.3& 1.1&0.16& 3.6&                    &                    &F&0\\
113&266.23050&-29.19856&    17.4&    15.1&15.9& 1.2&0.23& 3.8&                    &                    &F&0\\
114&266.15797&-29.16740&    11.8&    10.6&11.0& 1.0&0.09& 3.7&                    &                    &F&0\\
115&266.26565&-29.24893&        &        &16.2& 1.3&0.39& 4.5&                    &                    &F&0\\
116&266.29710&-29.26860&        &        &15.6& 1.2&0.22& 4.2&                    &                    &F&0\\
117&266.61068&-28.77401&    14.1&    13.0&13.4& 1.0&0.10& 4.1&                    &                    &F&0\\
118&266.63283&-28.77597&    11.2&    10.7&11.2& 1.0&0.10& 3.7&                    &                    &F&0\\
119&266.63634&-28.76893&    13.0&    12.0&12.5& 1.0&0.10& 4.1&                    &                    &F&0\\
120&266.58695&-28.78300&    14.1&    12.5&13.2& 1.0&0.09& 3.6&                    &                    &F&0\\
121&266.60519&-28.78461&    13.9&    13.7&13.7& 1.0&0.12& 4.5&                    &                    &F&0\\
122&266.60899&-28.81798&        &        &16.4& 2.0&1.04& 3.6&                    &                    &F&0\\
123&266.62481&-28.82561&    16.2&    14.7&15.1& 1.1&0.18& 3.7&                    &                    &F&0\\
124&266.55397&-28.78787&    14.6&    13.1&13.7& 1.1&0.14& 3.6&                    &                    &F&0\\
125&266.50629&-28.73107&    13.5&    11.7&12.5& 1.0&0.10& 3.5&                    &                    &F&0\\
126&266.50137&-28.75974&        &        &14.6& 1.1&0.13& 4.4&                    &                    &F&0\\
127&266.48354&-28.73089&        &        &14.1& 1.1&0.16& 4.3&                    &                    &F&0\\
128&266.47761&-28.77630&    15.2&    14.7&15.2& 1.1&0.14& 3.6&                    &                    &F&0\\
129&266.50941&-28.80310&        &        &16.1& 1.2&0.26& 3.8&                    &                    &F&0\\
130&266.55996&-28.83152&        &        &16.9& 5.5&4.53& 3.8&                    &                    &Q&0\\
131&266.57831&-28.83012&    12.5&    11.1&11.8& 1.1&0.11& 3.9&                    &                    &F&0\\
132&266.47769&-28.79581& 9.5$^a$& 7.9$^a$& 8.7& 1.0&0.12& 4.1&                    &                    &F&0\\
133&266.46318&-28.82305&    13.3&    11.7&12.4& 1.0&0.13& 3.6&     FA\_23,Blu01\_2&               O4-6I&F&0\\
134&266.45930&-28.82120&    13.1&    11.7&12.4& 1.0&0.10& 3.8&    FA\_22,Blu01\_27&               O4-6I&A&0\\
135&266.46112&-28.82337&        &        &13.1& 1.0&0.12& 4.1&              FA\_47&                    &A&0\\
136&266.50684&-28.85286&    16.1&    14.7&15.3& 1.2&0.24& 4.3&                    &                    &F&0\\
137&266.49781&-28.86983&        &        &16.2& 1.6&0.63& 4.1&                    &                    &F&0\\
138&266.51972&-28.92703&        &        &15.4& 1.2&0.22& 4.2&                    &                    &F&0\\
139&266.51018&-28.88714&        &        &13.9& 1.1&0.13& 3.7&                    &                    &F&0\\
140&266.50676&-28.91239&        &        &15.8& 1.4&0.50& 4.5&                    &                    &F&0\\
141&266.52558&-28.91151&        &        &14.0& 1.1&0.11& 3.6&                    &                    &F&0\\
142&266.45463&-28.89033&        &        &17.2& 1.8&0.93& 3.5&                    &                    &F&0\\
143&266.49235&-28.94004&        &        &15.9& 1.2&0.23& 3.6&                    &                    &F&0\\
144&266.54233&-28.96240&        &        &16.4& 1.3&0.38& 4.2&                    &                    &F&1\\
145&266.53275&-28.97921&    13.5&    12.7&13.0& 1.1&0.12& 3.9&                    &                    &F&0\\
146&266.53097&-28.98847&    15.1&    13.6&14.0& 1.1&0.16& 3.9&                    &                    &F&0\\
147&266.49729&-28.97560&        &        &15.4& 1.0&0.15& 3.5&                    &                    &F&0\\
148&266.47656&-28.94451& 6.9$^a$& 6.2$^a$& 6.4& 1.0&0.09& 3.6&                    &                    &F&0\\
149&266.35176&-28.86764&    10.2&     9.9&10.2& 1.0&0.09& 3.6&                    &                    &F&0\\
150&266.34827&-28.87596&    13.9&    13.5&13.8& 1.0&0.10& 3.8&                    &                    &F&0\\
151&266.35103&-28.88436&13.3$^a$&11.1$^a$&14.7& 1.0&0.12& 3.6&                    &                    &F&0\\
152&266.37455&-28.90802&    12.2&    10.8&11.3& 1.1&0.11& 4.2&                    &                    &F&0\\
153&266.39766&-28.92007&    15.2&    13.6&15.0& 1.0&0.13& 3.7&                    &                    &F&0\\
154&266.40925&-28.93099&10.2$^a$&10.0$^a$&10.1& 1.0&0.09& 3.9&                    &                    &F&0\\
155&266.47375&-28.97749&        &        &12.7& 1.0&0.10& 3.5&                    &                    &F&2\\
156&266.41055&-28.96760&        &        &16.4& 1.3&0.39& 3.7&                    &                    &F&1\\
157&266.38881&-28.94577&        &        &14.7& 1.1&0.15& 3.7&                    &                    &F&2\\
158&266.32543&-28.88945&    15.7&    14.0&14.9& 1.0&0.14& 3.7&                    &                    &F&0\\
159&266.35630&-28.90058&        &        &15.8& 1.3&0.34& 4.2&                    &                    &F&0\\
160&266.31731&-28.94053&        &        &16.3& 1.7&0.73& 4.1&                    &                    &F&0\\
161&266.36608&-28.96094&    16.6&    14.7&15.7& 1.1&0.21& 3.8&                    &                    &F&0\\
162&266.41716&-29.00766&        &        &10.9& 1.3&0.34& 4.4&                 E20&            Ofpe/WN9&C&0\\
163&266.41468&-29.00987&        &        &13.1& 1.1&0.21& 4.0&                 E79&            Ofpe/WN9&C&0\\
164&266.41688&-29.00749&    11.8&     9.6&10.7& 1.1&0.18& 3.6&                 E19&            Ofpe/WN9&C&0\\
165&266.41718&-29.00808& 9.7$^a$& 7.0$^a$&10.4& 1.2&0.23& 4.1&                 E23&            Ofpe/WN9&C&0\\
166&266.40589&-29.01854&        &        &15.8& 3.0&2.03& 4.4&                    &                    &F&0\\
167&266.46015&-29.03160&        &        &14.2& 1.1&0.14& 4.3&                    &                    &F&0\\
168&266.42288&-29.04927&    15.2&    13.9&14.4& 1.0&0.12& 4.1&                    &                    &F&0\\
169&266.36347&-29.00132&    13.0&    11.7&12.2& 1.1&0.13& 4.4&                    &                    &F&0\\
170&266.28998&-28.96300&    14.1&    14.3&13.9& 1.0&0.10& 3.7&                    &                    &F&0\\
171&266.37991&-29.07801&        &        &16.2& 1.2&0.28& 3.9&                    &                    &F&0\\
172&266.33615&-29.05066&        &        &14.4& 1.2&0.22& 4.4&                    &                    &F&0\\
173&266.33671&-29.10024&    14.7&    13.7&14.5& 1.1&0.13& 4.2&                    &                    &F&0\\
174&266.35621&-29.10382&        &        &15.6& 1.2&0.24& 3.6&                    &                    &F&0\\
175&266.32597&-29.11103&        &        &15.2& 1.1&0.19& 4.3&                    &                    &F&2\\
176&266.25701&-29.04719&        &        &15.0& 1.2&0.30& 3.7&                    &                    &F&0\\
177&266.25826&-29.06427&        &        &17.0& 1.6&0.71& 3.6&                    &                    &F&1\\
178&266.26467&-29.08358&    15.8&    14.3&14.9& 1.1&0.17& 3.6&                    &                    &F&0\\
179&266.32870&-29.16098&        &        &16.5& 1.4&0.46& 3.6&                    &                    &F&1\\
180&266.32652&-29.16999&        &        &15.1& 1.2&0.23& 4.4&                    &                    &F&0\\
181&266.31943&-29.18635&        &        &14.9& 1.1&0.16& 3.8&                    &                    &F&0\\
182&266.30597&-29.16803&    15.9&    14.2&16.2& 1.4&0.41& 4.2&                    &                    &F&0\\
183&266.26768&-29.15742&    15.3&    13.9&14.7& 1.1&0.15& 3.7&                    &                    &F&0\\
184&266.32193&-29.21338&        &        &15.7& 1.3&0.36& 3.7&                    &                    &F&0\\
185&266.21550&-29.18349&    13.1&    11.9&12.5& 1.1&0.14& 3.9&                    &                    &F&0\\
186&266.19144&-29.19229&        &        &14.3& 1.0&0.11& 3.5&                    &                    &F&0\\
187&266.24675&-29.23896&    16.0&    14.4&16.0& 1.3&0.36& 3.7&                    &                    &F&0\\
188&266.26815&-29.29092&        &        &15.7& 1.2&0.22& 3.7&                    &                    &F&0\\
189&266.22312&-29.27025&        &        &15.0& 1.1&0.16& 3.7&                    &                    &F&0\\
\enddata
 \label{t:pal_sou_second}
\tablecomments{Units of R.A. and Decl. are decimal degrees. H and K band magnitudes are mainly from the SIRIUS catalog~\citep{nis06}, while the superscript 'a' indicates the magnitudes are from the 2MASS catalog (see \S~\ref{s:dis}). The ground-based spetroscopically identified counterparts of our Pa$\alpha$ emitting sources and their types are from:~\citet{fig99a,fig02,lie09,mar08,blu01,cot99,hom03,mun06,mik06,mau07,mau10a,mau10b,mau10c}. The `Location' column divides the sources into four groups: the sources inside the three clusters ('Q': Quintuplet, 'A': Arches, 'C': Center) and field sources outside the clusters ('F', see \S~\ref{s:dis}). The `Problem Index' column definition is defined in \S~\ref{s:pal_emit}}
 \end{deluxetable}

\begin{deluxetable}{ccccccc}
  \tabletypesize{\scriptsize}
  \tablecolumns{7}
  \tablecaption{Comparison with previous spectroscopic identifications}
  \tablewidth{0pt}
  \tablehead{
  \colhead{Location} &
  \colhead{Age} &
  \colhead{$r_{c}(pc)$} &
  \colhead{O If} &
  \colhead{LBV} &
  \colhead{WN} &
  \colhead{WC} \\
  }
  \startdata
  Arches & 1-2 & 0.19 & 2/2 & 0/0 & 12/12 & 0/0  \\
  Quintuplet & 3-6 & 1 & 5/65 & 1/2& 6/6 & 7/13  \\
  Center & 3-7 & 0.23  & 4/9 & 0/0 & 7/10 & 4/13\\
  Field & & & 14/15 & 1/1& 6/6 & 8/8  \\
  \enddata
 \label{t:comparison}
\tablecomments{The number ratio of our detected Paschen-$\alpha$ sources (above the slash) to spectroscopically identified evolved massive stars of different 
types in various groups. The age and $r_c$ are from~\citet{fig99a}. $r_c$ is the average distance of the stars from the centroid of the clusters.}
 \end{deluxetable}

\begin{deluxetable}{cccccccc}
  \tabletypesize{\scriptsize}
  \tablecolumns{8}
  \tablecaption{Undetected known emission line stars}
  \tablewidth{0pt}
  \tablehead{
  \colhead{Source} &
  \colhead{Stellar} &
  \colhead{R.A.} &
  \colhead{Decl.} &
  \colhead{} & 
  \colhead{} & 
  \colhead{} &
  \colhead{} \\
  \colhead{IDs} &
  \colhead{type} &
  \colhead{(J2000.0)} &
  \colhead{(J2000.0)} &
  \colhead{$m_{F190N}$} &
  \colhead{r} &
  \colhead{r-$\bar{r}$} &
  \colhead{$N_{s}$} \\
  \colhead{(1)} &
  \colhead{(2)} &
  \colhead{(3)} &
  \colhead{(4)}&
  \colhead{(5)}&
  \colhead{(6)} &
  \colhead{(7)} & 
  \colhead{(8)}\\
  }
  \startdata
E31&       WC9&266.41636&-29.00745&12.2&0.90&-0.05&-0.7\\
E32&     WC8/9&266.41743&-29.00813&12.1&0.92&-0.02&-0.2\\
E35&     WC8/9&266.41652&-29.00725&13.2&1.03& 0.09& 1.0\\
E41&  Ofpe/WN9&266.41705&-29.00868&10.9&1.20& 0.24& 3.2\\
E58&     WC5/6&266.41612&-29.00677&13.5&1.01& 0.09& 2.0\\
E59&       WC9&266.41779&-29.00686&13.9&1.51& 0.59& 2.8\\
E60&      WN7?&266.41545&-29.00826&13.5&1.42& 0.47& 2.3\\
E61&       WN7&266.41565&-29.00705&14.2&1.22& 0.31& 2.6\\
E70&  Ofpe/WN9&266.41827&-29.00644&13.5&1.63& 0.70& 2.1\\
E72&      WC9?&266.41904&-29.00786&12.0&0.89&-0.04&-0.6\\
E76&       WC9&266.41820&-29.01004&13.9&1.52& 0.60& 1.4\\
E80&       WC9&266.41862&-29.01009&12.7&1.36& 0.43& 3.5\\
 \enddata
 \label{t:pal_miss}
\tablecomments{The evolved massive stars within the Central cluster missed 
by our method. The source IDs are from~\citet{pau06}. Units of R.A. and Decl. 
are decimal degrees.}  
 \end{deluxetable}

\begin{figure*}[!thb]
  \centerline{
       \epsfig{figure=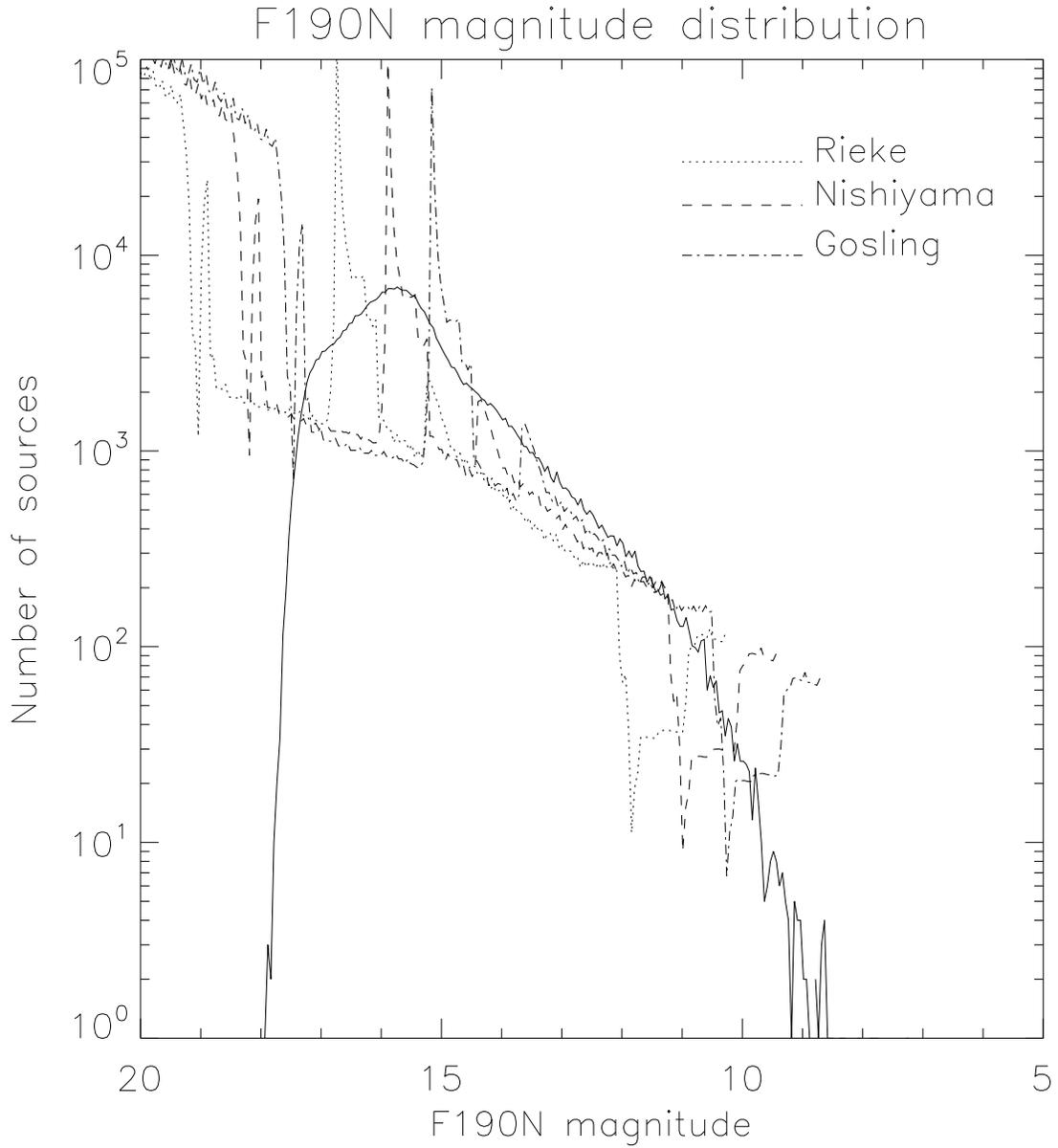,width=1\textwidth,angle=90}
       }
 \caption{Comparison of the magnitude distribution from our survey
   and those predicted by the Padova stellar evolutionary tracks of
   a stellar population with 2 Gyr old and solar metallicity, after
   corrected for the distance and extinction modulus. Assuming three extinction laws
(dotted: Rieke 1999, dashed: Nishiyama et al,
   2009 and dot-dashed: Gosling et al, 2009). }
 \label{f:lum_com}
 \end{figure*}

\begin{figure*}[!thb]
  \centerline{
       \epsfig{figure=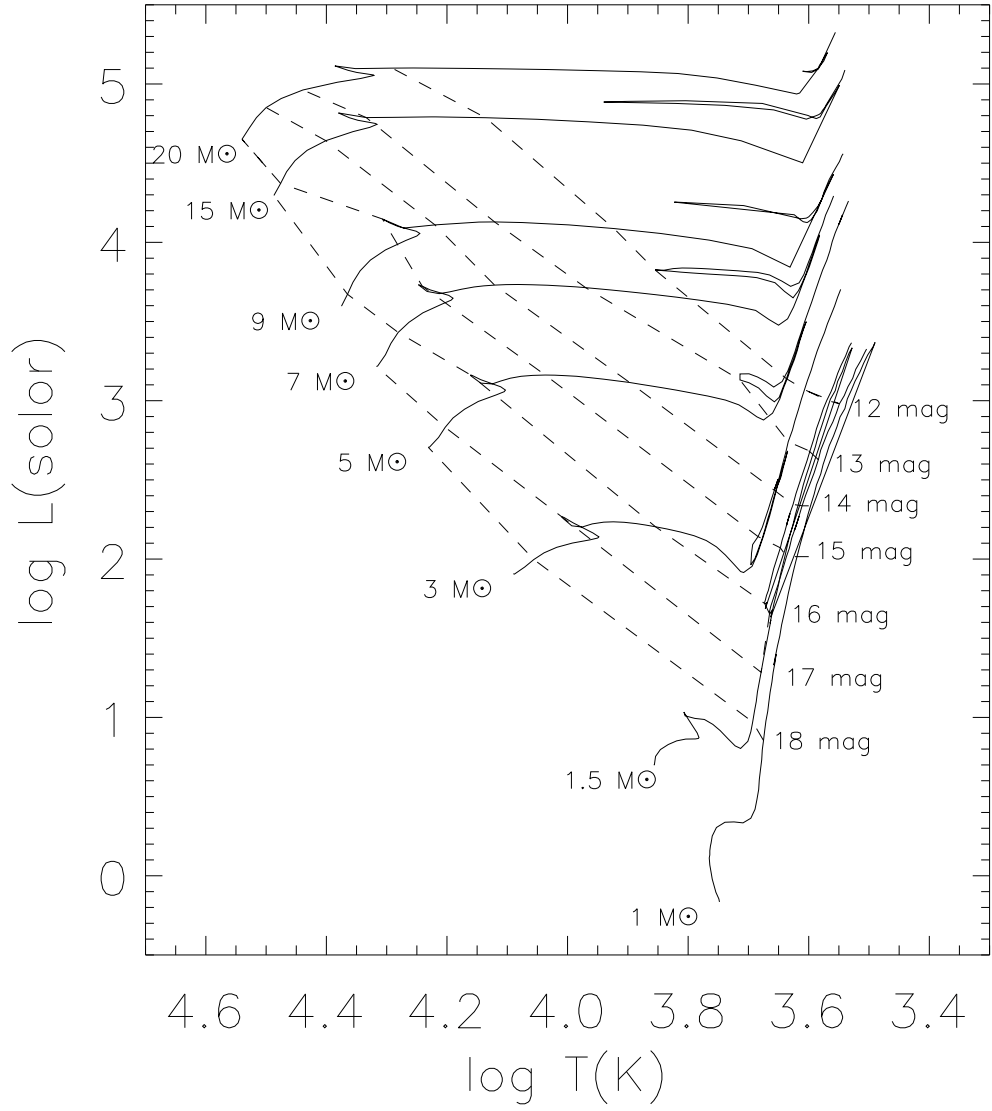,width=1.0\textwidth,angle=0}
       }
 \caption{1.90 $\mu$m magnitude (black dashed lines) contours overlaid on the Padova model
stellar evolutionary tracks (black solid lines). 
}
 \label{f:det_limit}
 \end{figure*}

\end{document}